\documentclass[twocolumn,noshowpacs,showkeys,amsmath,amssymb,preprintnumbers,nofootinbib,floatfix,rmp]{revtex4}

\usepackage[usenames]{color}
\usepackage{graphicx}
\usepackage{dcolumn}
\usepackage{bm}
\usepackage{bbm}

\usepackage{mathrsfs}

\usepackage{verbatim}
\usepackage{amsmath}
\usepackage{fancyhdr} 
\usepackage[toc,page]{appendix}

\begin{document}


\title{Einstein's vierbein field theory of curved space}

\author{Jeffrey Yepez}

\date{January 20, 2008}


\address{Air Force Research Laboratory, Hanscom Air Force Base, MA  01731}

\begin{abstract}
General Relativity theory is reviewed following the vierbein field theory approach proposed in 1928 by Einstein.  It is based on the vierbein field taken as the ``square root'' of the metric tensor field.  
 Einstein's vierbein theory is a gauge field theory for gravity;  the vierbein field playing the role of a gauge field but not exactly like the vector potential field does in Yang-Mills theory--the correction to the derivative (the covariant derivative) is not proportional to the vierbein field as it would be if gravity were strictly a Yang-Mills theory.  Einstein discovered the spin connection in terms of the vierbein fields to take the place of the conventional affine connection.
  To date, one of the most important applications of the vierbein representation is for the derivation of the correction to a 4-spinor quantum field transported in curved space, yielding the correct form of the covariant derivative.  Thus, the vierbein field theory is the most natural way  to represent a relativistic quantum field theory in curved space.  Using the vierbein field theory, presented is a derivation of the the Einstein equation and then the Dirac equation in curved space.  Einstein's original 1928 manuscripts translated into English are  included.

\end{abstract}


\maketitle

\definecolor{red_ud}{rgb}{1,0,0} 
\definecolor{red_cs}{rgb}{.75,0,0} 
\definecolor{red_tb}{rgb}{.5,0,0} 

\definecolor{blue_ud}{rgb}{0,0,1} 
\definecolor{blue_cs}{rgb}{0,0,.75} 
\definecolor{blue_tb}{rgb}{0,0,.5} 

\definecolor{green_ud}{rgb}{0,1,0} 
\definecolor{green_cs}{rgb}{0,.75,0} 
\definecolor{green_tb}{rgb}{0,.5,0} 

\tableofcontents

\newpage
\section{Introduction}

The purpose of this manuscript is to provide a self-contained review of the procedure for deriving the Einstein equations for gravity and the Dirac equation in curved space using the vierbein field theory.   This gauge field theory approach to General Relativity (GR) was discovered by Einstein in 1928 in his pursuit of  a unified field theory of gravity and electricity.  He originally published this approach in two successive letters appearing one week apart \cite{einstein-1928a,einstein-1928b}.  The first manuscript, a seminal contribution to mathematical physics, adds the concept of distant parallelism to Riemann's theory of curved manifolds that is based on comparison of distant vector magnitudes, which before Einstein did not incorporate comparison of distant directions.


Historically there appears to have been a lack of interest in Einstein's research following his discovery of general relativity, principally from the late 1920's onward. 
Initial enthusiasm for Einstein's unification approach turned into a general rejection.
In Born's July 15th, 1925 letter (the most important in the collection) to Einstein following the appearance of his student Heisenberg's revolutionary paper on the matrix representation of quantum mechanics, Born writes \cite{Born_Einstein_letters}:
\begin{quote}
Einstein's field theory \dots was intended to unify electrodynamics and gravitation \dots.
I think that my enthusiam about the success of Einstein's idea was quite genuine.  In those days we all thought that his objective, which he pursued right to the end of his life, was attainable and also very important.  Many of us became more doubtful when other types of fields emerged in physics, in addition to these; the first was Yukawa's meson field, which is a direct generalization of the electromagnetic field and describes nuclear forces, and then there were the fields which belong to the other elementary particles.  After that we were inclined to regard Einstein's ceaseless efforts as a tragic error.
\end{quote}
Weyl and Pauli's rejection of Einstein's thesis of distant parallelism 
also helped paved the way for the view that Einstein's findings had gone awry.
Furthermore, as the belief in the fundamental correctness of quantum theory solidified by burgeoning experimental verifications, 
the theoretical physics community seemed more inclined to latch onto Einstein's purported repudiation of quantum mechanics: he failed to grasp the most important direction of twentieth century physics.

Einstein announced his goal of achieving a unified field theory before he published firm results.  It is already hard not to look askance at an audacious unification agenda, but it did not help 
 when the published version of the manuscript had a fundamental error in its opening equation;  even though  this error was perhaps introduced by the publisher's typist, it can cause confusion.\footnote{The opening equation (1a) was originally typeset as \[ \mathfrak{H} = h\, g^{\mu\nu}\; , \; {\Lambda_\mu}^\alpha_\beta, \; {\Lambda_\nu}^\beta_\alpha, \;\cdots\] offered as the Hamiltonian whose variation at the end of the day yields the Einstein and Maxwell's equations.  I have corrected this in the translated manuscript in the appendix.}
  As far as I know at the time of this writing in 2008, the two  1928 manuscripts have never been translated into English.   English versions of these manuscripts are provided as part of this review--see the Appendix--and are included to contribute to its completeness.


In the beginning of the year 1928, Dirac introduced his famous square root of the Klein-Gordon equation, establishing the starting point for the story of relativistic quantum field theory, in his paper on the quantum theory of the electron \cite{1928RSPSA.117..610D}.  This groundbreaking paper by Dirac may have inspired Einstein, who completed his manuscripts a half year later in the summer of 1928.  With deep insight, Einstein introduced the vierbein field, which constitutes the square root of the metric tensor field.\footnote{The culmination of Einstein's new field theory approach appeared in Physical Review in 1948 \cite{RevModPhys.20.35}, entitled ``A Generalized Theory of Gravitation."}  
 Einstein and Dirac's square root theories  mathematically fit well together;  they even become joined at the hip when one considers the dynamical behavior of chiral matter in curved space.

 Einstein's second manuscript represents a simple and intuitive attempt to unify gravity and electromagnetism.  
 He originally developed the vierbein field theory approach with the goal of unifying gravity and quantum theory, a goal which he never fully attained with this representation.  
 Nevertheless, the vierbein field theory approach represents progress in the right direction.   Einstein's unification of gravity and electromagnetism, using only fields in four-dimensional spacetime, is conceptually much simpler than the well known Kaluza-Klein approach to unification that posits an extra compactified spatial dimension.  But historically it was the Kaluza-Klein notion of extra dimensions that gained popularity as it was generalized to string theory.  In contradistinction, Einstein's approach requires no extra constructs, just the intuitive notion of distant parallelism. In the Einstein vierbein field formulation of the connection and curvature, the basis vectors in the tangent space of a spacetime manifold are not derived from any coordinate system of that manifold.


Although Einstein is considered one of the founding fathers of quantum mechanics, he is not presently considered one of the founding fathers of relativistic quantum field theory in flat space.  This is understandable since in his theoretical attempts to discover a unified field theory he did not predict any of the new leptons or quarks, nor their weak or strong gauge interactions, in the Standard Model of particle physics that emerged some two decades following his passing.    However, Einstein did employ local rotational invariance as the gauge symmetry in the first 1928 manuscript and discovered what today we call the {\it spin connection}, the gravitational gauge field associated with the Lorentz group as the local symmetry group ({\it viz.} local rotations and boosts).  

This he accomplished about three decades before Yang and Mills discovered nonabelian gauge theory \cite{PhysRev.96.191},  the antecedent to the Glashow-Salam-Weinberg electroweak unification theory \cite{Glashow1961579,Salam1966683,PhysRevLett.19.1264} that is the cornerstone of the Standard Model.  
Had Einstein's work toward unification been more widely circulated instead of rejected, perhaps Einstein's original discovery of $n$-component gauge field theory would be broadly considered the forefather of Yang-Mills theory.\footnote{Einstein treated the general case of an $n$-Bein field.}
In Section~\ref{Similarity_to_Yang_Mills_gauge_theory}, I sketch a few of the strikingly similarities between the vierbein field representation of gravity and the Yang-Mills nonabelian gauge theory.
%


With the hindsight of 80 years  of theoretical physics development from the time of the first appearance of Einstein's 1928 manuscripts, today one can see the historical rejection is a mistake.
Einstein could rightly be considered one of the founding fathers of relativistic quantum field theory in curved space, and these 1928 manuscripts 
should not be  forgotten. Previous attempts have been made to revive interest in the vierbein theory of gravitation purely on the grounds of their superior use for representing GR, regardless of unification \cite{PhysRev.165.1420}. Yet, this does not go far enough.  One should also make the case for the requisite application to quantum fields in curved space.  

Einstein is famous for (inadvertently) establishing another field of contemporary physics with the discovery of distant quantum entanglement.  Nascent quantum information theory was borne from the seminal 1935  Einstein, Podolsky, and Rosen (EPR) Physical Review paper\footnote{This is the most cited physics paper ever and thus a singular exception to the general lack of interest in Einstein's late research.},  ``Can Quantum-Mechanical Description of Physical Reality Be Considered Complete?"  
Can Einstein equally be credited for establishing the field of quantum gravity, posthumously?


The concepts of vierbein field theory are simple, and the mathematical development is straightforward, but in the literature one can find the vierbein theory a notational nightmare, making this pathway to GR appear more difficult than it really is, and hence less accessible.   In this manuscript, I hope to offer an accessible pathway to GR and the Dirac equation in curved space.
The development of the vierbein field theory presented here borrows first from treatments given by Einstein himself  that I have discussed above \cite{einstein-1928a,einstein-1928b}, as well as excellent treatments by Weinberg \cite{Weinberg_72} and Carroll \cite{Carroll_2004}.\footnote{Both Weinberg and Carroll treat GR in a traditional manner.  Their respective explanations of Einstein's vierbein field theory are basically incidental.  Carroll relegates his treatment to a single appendix.} \footnote{Another introduction to GR but which does not deal with the vierbein theory extensively is the treatment by D'Inverno \cite{DInverno_95}.} However, Weinberg and Carroll review vierbein theory as a sideline to their main approach to GR, which is the standard coordinate-based approach of differential geometry.  An excellent treatment of quantum field theory in curved space is given by Birrell and Davies \cite{Birrell_Davies_82}, but again with a very brief description of the vierbein representation of GR.  Therefore, it is hoped that a self-contained review of Einstein's vierbein theory and the associated formulation of the relativistic wave equation should be helpful gathered together in one place.

\subsection{Similarity to Yang-Mills gauge theory}
\label{Similarity_to_Yang_Mills_gauge_theory}

This section is meant to be an outline comparing the structure of GR  and Yang-Mills (YM) theories \cite{PhysRev.96.191}.  There  are many previous treatments of this  comparison---a recent treatment by Jackiw is recommended \cite{jackiw_2005}. The actual formal review of the vierbein theory does not begin until Section~\ref{vierbein_mathematical_framework}.

The dynamics of the metric tensor field in GR can be cast in the form of a YM gauge theory used to describe the dynamics of the quantum field in the Standard Model. 
In GR, dynamics is invariant under an external local transformation, say $\Lambda$, of the Lorentz group SO(3,1) that includes rotations and boosts. Furthermore, any quantum dynamics occurring within the spacetime manifold is invariant under internal local Lorentz transformations, say $U_\Lambda$, of the spinor representation of the SU(4) group.  Explicitly, the internal Lorentz transformation of a quantum spinor field in unitary form is
\begin{equation}
\label{qiGR_internal_unitary_Lorentz_transformation}
U_\Lambda = e^{-\frac{i}{2} \omega_{\mu \nu}(x) S^{\mu \nu}},
\end{equation}
where $S^{\mu \nu}$ is the tensor generator of the transformation.\footnote{A $4\times 4$ fundamental representation of SU(4) are the $4^2-1=15$ Dirac matrices, which includes four vectors $\gamma^\mu$, six tensors $\sigma^{\mu\nu}=\frac{i}{2}[\gamma^\mu, \gamma^\nu]$, one pseudo scalar $i\gamma^0\gamma^1\gamma^2\gamma^3\equiv \gamma^5$, and four axial vectors $\gamma^5\gamma^\mu$.  The generator associated with the internal Lorentz transformation is  $S^{\mu \nu}=\frac{1}{2}\sigma^{\mu\nu}$.} 
Similarly, in the Standard Model, the dynamics of  the Dirac particles (leptons and quarks) is invariant under local transformations of the internal gauge group, SU(3) for color dynamics and SU(2) for electroweak dynamics.\footnote{A $3\times 3$ fundamental representation of SU(3) are the $3^2-1=8$ Gell-Mann matrices.  And, a $2\times 2$ fundamental representation of SU(2) are the $2^2-1=3$ Pauli matrices.}
 The internal unitary transformation of the multiple component YM field in unitary form is 
\begin{equation}
\label{YM_gauge_transformation}
 U=e^{i \Theta^a(x)T^a}, 
\end{equation}
where the hermitian generators  $T^a = T^{a\dagger}$ are in the adjoint representation of the gauge group. The common unitarity of (\ref{qiGR_internal_unitary_Lorentz_transformation}) and (\ref{YM_gauge_transformation}) naturally leads to many parallels between the GR and YM theories.

From (\ref{qiGR_internal_unitary_Lorentz_transformation}), it seems that $\omega_{\mu\nu}(x)$ should take the place of the gauge potential field, and in this case it is clearly a second rank field quantity.   But in Einstein's action in his vierbein field representation of GR, the lowest order fluctuation of a second-rank vierbein field itself
\begin{equation*}
{e^\mu}_a(x)=\delta^\mu_a + {k^\mu}_a(x) + \cdots
\end{equation*}
  plays the roles of the potential field where  the gravitational field strength ${F^{\mu\nu}}_a$ is related to a quantity of the form
\begin{equation}
\label{gravitational_field_strength}
{F^{\mu\nu}}_a=\partial^\mu {k^\nu}_a(x) - \partial^\nu {k^\mu}_a(x).
\end{equation}
(\ref{gravitational_field_strength}) vanishes for  apparent or pseudo-gravitational fields that occur in rotating or accelerating local inertial frames but does not vanish for gravitational fields associated with curved space.   Einstein's expression for the Lagrangian density (that he presented as a footnote in his second manuscript) gives rise to a field strength of the form of (\ref{gravitational_field_strength}).  An expanded version of his proof of the equivalence of his vierbein-based action principle with gravity in the weak field limit is presented in Section~\ref{Einsteins_action_in_vierbein_field_variables}.

In any case, a correction  to the usual derivative \[\partial_\mu \rightarrow {\cal D}_\mu\equiv \partial_\mu + \Gamma_\mu\] 
is necessary in the presence of a gravitational field.  The local transformation has the form
\begin{subequations}
\label{qiGR_Gamma_pseudo_gauge_transformation}
\begin{eqnarray}
\psi 
&\rightarrow&
U_\Lambda\psi
\\
\Gamma_\mu
&\rightarrow&
U_\Lambda
\Gamma_\mu
U^\dagger_\Lambda
-
\left(
\partial_\mu
U_\Lambda
\right)
U^\dagger_\Lambda.
\end{eqnarray}
\end{subequations}
(\ref{qiGR_Gamma_pseudo_gauge_transformation}) is derived in Section~\ref{Invariance_in_curved_space}. This is similar to Yang-Mills gauge theory where a correction  to the usual derivative \[\partial_\mu \rightarrow {\cal D}_\mu\equiv \partial_\mu -i  A_\mu\] 
is also necessary in the presence of a non-vanishing gauge field. In YM, the gauge transformation has the form
\begin{subequations}
\begin{eqnarray}
\psi&\rightarrow &U\,\psi\\
A_{\mu}&\rightarrow& UA_{\mu}U^{\dagger} - \,U\partial_{\mu}U^{\dagger},
\end{eqnarray}
 \end{subequations}
 which is just like (\ref{qiGR_Gamma_pseudo_gauge_transformation}).
Hence, this ``gauge field theory'' approach to GR is useful in deriving the form of the Dirac equation in curved space.  In this context, the requirement of invariance of the relativistic quantum wave equation to local Lorentz transformations leads to a correction of the form
\begin{subequations}
\begin{eqnarray}
\Gamma_{\mu}
&=&
 \frac{1}{2} {e^\beta}_k
\left(
\partial_\mu e_{\beta h}
\right)
S^{h k}
\\
&=&
\partial_\mu
\left(
 \frac{1}{2} 
 k_{\beta h}
S^{h \beta}
\right)
+\cdots
\end{eqnarray}
 \end{subequations}
This is derived in Section~\ref{relativistic_chiral_matter_in_curved_space}.
A problem that is commonly cited regarding the gauge theory representation of General Relativity 
is that the correction is not directly proportional to the gauge potential  as would be the case if it were strictly a YM theory ({\it e.g.} we should be able to write $\Gamma_\mu(x) = u^a {k_{\mu a}}(x)$ where $u^a$ is some constant four-vector).

\section{Mathematical framework}
\label{vierbein_mathematical_framework}

\subsection{Local basis}

The conventional coordinate-based approach to GR uses a ``natural'' differential basis for the tangent space $T_p$ at a point $p$ given by the partial derivatives of the coordinates at $p$
\begin{equation}
\label{contravariant_othronormal_basis_coordinate_vector_component}
\hat {\bf e}_{(\mu)} =\partial_{(\mu)}.
\end{equation}
Some 4-vector $A \in T_p$ has components
\begin{equation}
A = A^\mu \hat {\bf e}_{(\mu)}  =(A_0,A_1,A_2,A_3).
\end{equation}
To help reinforce the construction of the frame, we use a triply redundant notation of using a bold face symbol to denote a basis vector $\bf e$,  applying a caret symbol $\,\,\hat{\bf e}\,\,$ as a hat to denote a unit basis vector, and enclosing the component subscript with parentheses $\hat{\bf e}_{(\mu)}$ to denote a component of a basis vector.\footnote{With $\hat {\bf e}_{(\bullet)} \in \{e_0,e_1,e_2,e_3\}$ and $\partial_{(\bullet)} \in \{\partial_0,\partial_1,\partial_2,\partial_3\}$, some authors write  (\ref{contravariant_othronormal_basis_coordinate_vector_component}) concisely as 
\begin{equation*}
\label{contravariant_othronormal_basis_coordinate_vector_component_2}
e_\mu = \partial_\mu.
\end{equation*}
I will not use this notation because I would like to reserve $e_\mu$ to represent the lattice vectors $e_\mu\equiv\gamma_a {e_\mu}^a$ where $\gamma_a$ are Dirac matrices and  ${e_\mu}^a$ is the vierbein field defined below.}
 It should be nearly impossible to confuse a component of an orthonormal basis vector in $T_p$ with a component of any other type of object.
Also, the use of a Greek index, such as $\mu$, denotes a component in a coordinate system representation.  Furthermore, the choice of writing the component index as a superscript as in $A^\mu$
is the usual convention for indicating this is a component of a contravariant vector.  A contravariant vector is often just called a vector, for simplicity of terminology. 

In the natural differential basis, the cotangent space, here denoted by $T^\ast_p$, is spanned by the differential elements 
\begin{equation}
\label{covariant_othronormal_basis_coordinate_vector_component}
\hat {\bf e}^{(\mu)} ={\bf dx}^{(\mu)},
\end{equation}
which lie in the direction of the gradient of the coordinate functions.\footnote{
Again, with $\hat {\bf e}^{(\bullet)} \in \{e^0,e^1,e^2,e^3\}$ and ${\bf dx}^{(\bullet)} \in \{dx^0,dx^1,dx^2,dx^3\}$, for brevity some authors write (\ref{covariant_othronormal_basis_coordinate_vector_component}) as 
\begin{equation*}
\label{covariant_othronormal_basis_coordinate_vector_component_2}
e^\mu = dx^\mu.
\end{equation*}
But I reserve $e^\mu$ to represent anti-commuting 4-vectors, $e^\mu\equiv\gamma^a {e^\mu}_a$, unless otherwise noted.
}
$T^\ast_p$ is also called the dual space of $T_p$.
Some dual 4-vector $A \in T_p^\ast$ has components
\begin{equation}
A = A_\mu \hat {\bf e}^{(\mu)} =  g_{\mu\nu} A^\nu {\bf e}^{(\mu)}.
\end{equation}
Writing the component index $\mu$ as a subscript in $A_\mu$
again follows the usual convention for indicating one is dealing with a component of a covariant vector.  Again, in an attempt to simplify terminology, a covariant vector is often called a 1-form, or simply a dual vector.   Yet, remembering that a vector and dual vector (1-form) refer to an element of the tangent space $T_p$ and the cotangent space $T^\ast_p$, respectively, may not seem all that much easier than remembering the prefixes contravariant and covariant in the first place.   

The dimension of (\ref{contravariant_othronormal_basis_coordinate_vector_component}) is inverse length, $[\hat {\bf e}_{(\mu)}]=\frac{1}{L}$, and this is easy to remember because a first derivative of a function is always tangent to that function.  For a basis element, $\mu$ is a subscript when $L$ is in the denominator. Then (\ref{covariant_othronormal_basis_coordinate_vector_component}), which lives in the cotangent space and as the dimensional inverse of  (\ref{contravariant_othronormal_basis_coordinate_vector_component}), must have dimensions of length $[\hat {\bf e}^{(\mu)}]=L$.   So, for a dual basis element, $\mu$ is a superscript when $L$ is in the numerator. That they are dimensional inverses is expressed in the following tensor product space
\begin{equation}
\hat {\bf e}^{(\mu)}\otimes \hat {\bf e}_{(\nu)} ={\mathbf{1}^\mu}_\nu,
\end{equation}
where $\mathbf{1}$ is the identity, which is of course dimensionless.

We are free to choose any orthonormal basis we like to span $T_p$, so long as it has the appropriate signature of the manifold on which we are working.  To that end, we introduce a set of basis vectors $\hat{\bf e}_{(a)}$, which we choose as non-coordinate unit vectors, and we denote this choice by using small Latin letters for indices of the non-coordinate frame. 
With this understanding, the inner product may be expressed as
\begin{equation}
\left(
\hat{\bf e}_{(a)},
\hat{\bf e}_{(b)}
\right)
=
\eta_{a b},
\end{equation}
where $\eta_{a b} = \text{diag}(1, -1, -1, -1)$ is the Minkowski metric of flat spacetime.

\subsection{Vierbein field}

This orthonormal basis that is independent of the coordinates is termed a {\it tetrad basis}.\footnote{To help avoid confusion, please note that the term {\it tetrad} in the literature is often used as a synonym for the term {\it vierbein}.  Here we use the terms to mean two distinct objects: $\hat{\bf e}_{(a)}$ and $e^\mu(x)$, respectively.} 
Although we cannot find a coordinate chart that covers the entire curved manifold, we can choose a fixed orthonormal basis that is independent of position.  
Then, from a local perspective, any vector can be expressed as a linear combination of the fixed tetrad basis vectors at that point. 
Denoting an element of the tetrad basis by $\hat{\bf e}_{(a)}$, we can express the coordinate basis (whose value depends on the local curvature at a point $x$ in the manifold) in terms of the tetrads as the following linear combination
\begin{equation}
\label{tetrad_basis_to_coordinate_basis_transformation}
\hat{\bf e}_{(\mu)}(x)
=
{e_\mu}^a(x) \,\hat{\bf e}_{(a)},
\end{equation}
where the functional components ${e_\mu}^a(x)$ form a $4 \times 4$ invertible matrix.  We will try not to blur the distinction between a vector and its components.  The term ``vierbein field'' is used to refer to the whole transformation matrix in (\ref{tetrad_basis_to_coordinate_basis_transformation}) with 16 components, denoted by ${e_\mu}^a(x)$.  The vierbeins ${e_\mu}^a(x)$, for $a=1,2,3,4$, comprise four legs--{\it vierbein} in German means four-legs.

The inverse of the vierbein has components, ${e^\mu}_a$ (switched indices), that satisfy the  orthonormality conditions
\begin{equation}
\label{vierbein_orthonormality_conditions}
{e^\mu}_a(x) {e_\nu}^a(x)
=
\delta^\mu_\nu,
\quad
{e_\mu}^a(x) {e^\mu}_b(x)
=
\delta^a_b.
\end{equation}
The inverse vierbein serves as a transformation matrix that allows one to represent the tetrad basis $\hat{\bf  e}_{(a)}(x)$ in terms of the coordinate basis $\hat{\bf  e}_{(\mu)}$:
\begin{equation}
\hat{\bf  e}_{(a)}
=
{e^\mu}_a(x) \,\hat{\bf  e}_{(\mu)}.
\end{equation}
Employing the metric tensor $g_{\mu \nu}$ to induce the product of the vierbein field and inverse vierbein field, the inner product-signature constraint is
\begin{equation}
g_{\mu \nu}(x) {e^\mu}_a(x)\, {e^\nu}_b(x)
=
\eta_{a b},
\end{equation}
or using (\ref{vierbein_orthonormality_conditions}) equivalently we have
\begin{equation}
g_{\mu \nu}(x) = {e_\mu}^a(x) {e_\nu}^b(x) \eta_{a b}.
\end{equation}
So, the vierbein field is the ``square root'' of the metric.

Hopefully, you can already see why one should include the vierbein field theory as a member of our tribe of ``square root'' theories.  These include the Pythagorean theorem for the distance interval $ds =\sqrt{\eta_{\mu\nu} dx^\mu dx^\nu}= \sqrt{dt^2 - dx^2 - dy^2 - dz^2}$, the mathematicians' beloved complex analysis (based on $\sqrt{-1}$ as the imaginary number),  quantum mechanics (e.g. pathways are assigned amplitudes which are the square root of probabilities), quantum field theory (e.g. Dirac equation as the square root of the Klein Gordon equation), and quantum computation based on the universal $\sqrt{\text{\sc swap}}$ conservative quantum logic gate.  To this august list we add the vierbein as the square root of the metric tensor.

Now, we may form a dual orthonormal basis, which we denote by $\hat{\bf e}^{(a)}$ with a Latin superscript, of 1-forms in the cotangent space $T^\ast_p$ that satisfies the tensor product condition
\begin{equation}
\hat{\bf e}^{(a)} \otimes
\hat{\bf e}_{(b)}
=
{\mathbf{1}^a}_b.
\end{equation}
This non-coordinate basis 1-form can be expressed as a linear combination of coordinate basis 1-forms
\begin{equation}
\label{noncoordinate_basis_1_form}
\hat{\bf e}^{(a)}
=
{e_\mu}^a(x) \, \hat{\bf e}^{(\mu)}(x),
\end{equation}
where $\hat{\bf e}^{(\mu)} 
=
dx^\mu$,
and vice versa using the inverse vierbein field
\begin{equation}
\hat{\bf e}^{(\mu)}(x)
=
{e^\mu}_a(x)\, \hat{\bf e}^{(a)}.
\end{equation}

Any vector at a spacetime point has components in the coordinate and non-coordinate orthonormal basis
\begin{equation}
{\bf V} =
V^\mu \,\hat{\bf e}_{(\mu)}
=
V^a \,\hat{\bf e}_{(a)}.
\end{equation}
So, its components are related by the vierbein field transformation
\begin{equation}
V^a = {e_\mu}^a V^\mu\qquad\text{and}\qquad V^\mu = {e^\mu}_a V^a.
\end{equation}
The vierbeins allow us to switch back and forth between Latin and Greek bases.

Multi-index tensors can be cast in terms of mixed-index components, as for example
\begin{equation}
{V^a}_b
=
{e_\mu}^a {V^\mu}_b
=
{e^\nu}_b {V^a}_\nu
=
{e_\mu}^a {e^\nu}_b {V^\mu}_\nu.
\end{equation}
The behavior of inverse vierbeins is consistent with the conventional notion of raising and lowering indices.   Here is an example with the metric tensor field and the Minkowski metric tensor
\begin{equation}
{e^\mu}_a
=
g^{\mu \nu} \eta_{a b} \,{e_\nu}^b.
\end{equation}
The identity map has the form
\begin{equation}
{\bf e} = {e_\nu}^a {\bf dx}^{(\nu)}
\otimes \hat{\bf e}_{(a)}.
\end{equation}

We can interpret ${e_\nu}^a$ as a set of four Lorentz 4-vectors.  That is, there exists one 4-vector for each non-coordinate index $a$. 

We can make local Lorentz transformations (LLT) at any point.  The signature of the Minkowski metric is preserved by a Lorentz transformation
\begin{equation}
\text{LLT:}\quad
\hat {\bf e}_{(a)} 
\rightarrow
\hat {\bf e}_{(a')}
=
{\Lambda^a}_{a'}(x) \hat {\bf e}_{(a)},
\end{equation}
where ${\Lambda^{a}}_{a'}(x)$ is an inhomogeneous ({\it i.e.} position dependent) transformation that satisfies
\begin{equation}
{\Lambda^a}_{a'} {\Lambda^b}_{b'} \eta_{a b}
=
\eta_{a' b'}.
\end{equation}
A Lorentz transformation can also operate on basis 1-forms, in contradistinction to the ordinary Lorenz transformation ${\Lambda^{a'}}_{a}$ that operates on basis vectors.  ${\Lambda^{a'}}_{a}$ transforms upper (contravariant) indices, while ${\Lambda^a}_{a'}$ transforms lower (covariant) indices.

And, we can make general coordinate transformations (GCT)
\begin{equation}
\text{GCT:}\quad {T^{a \mu}}_{b \nu}\rightarrow
{T^{a' \mu'}}_{b' \nu'}
=
\underbrace{{\Lambda^{a'}}_a}_{
\begin{matrix}
      \text{\tiny prime}    \\
      \text{\tiny 1st}  \\
      \text{\tiny (contra-}\\
      \text{\tiny variant)}
\end{matrix}}
\frac{\partial x^{\mu'}}{\partial x^\mu}
\underbrace{{\Lambda^b}_{b'}}_{\begin{matrix}
      \text{\tiny prime}    \\
      \text{\tiny 2nd}  \\
      \text{\tiny (co-}\\
      \text{\tiny variant)}
\end{matrix}}
\frac{\partial x^{\nu}}{\partial x^{\nu'}}
{T^{a \mu}}_{b \nu}.
\end{equation}

\section{Connections}

\subsection{Affine connection}

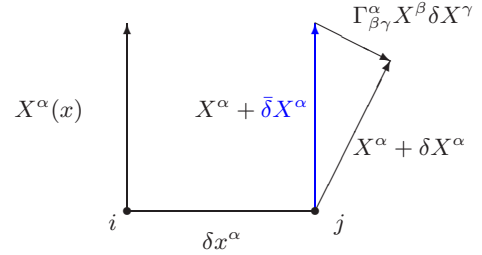
\begin{figure}[htbp]
\begin{center}
\setlength{\unitlength}{.5mm}
\begin{picture}(65,80)


\put(0,0){\vector(0,1){50}}

\put(0,0){\line(1,0){50}}

\put(50,0){\color{blue}\vector(0,1){50}}

\put(50,0){\vector(1,2){20}}

\put(50,50){\vector(2,-1){20}}

\put(-5,-5){$i$}
\put(55,-5){$j$}

\put(20,-10){$\delta x^\alpha$}
\put(18,25){$X^\alpha+\color{blue}\bar \delta X^\alpha$}
\put(-30,25){$X^\alpha(x)$}
\put(60,15){$X^\alpha+ \delta X^\alpha$}
\put(60,50){$\Gamma^\alpha_{\beta\gamma}X^\beta\delta X^\gamma$}

\put(50,0){\circle*{2}}
\put(0,0){\circle*{2}}

\end{picture}
\end{center}
\caption{\label{parallel_transport}\footnotesize  Two  spacetime points $x^\alpha$ and $x^\alpha + \delta x^\alpha$, labeled as $i$ and $j$, respectively.  The 4-vector at point $i$ is $X^\alpha(x)$, and the 4-vector at nearby point $j$ is $X^\alpha(x+\delta x)=X^\alpha(x)+ \delta X^\alpha(x)$.  The parallel transported 4-vector at $j$ is $ X^\alpha(x)+{\color{blue}\bar \delta X^\alpha}(x)$ ({\color{blue}blue}).  The affine connection is $\Gamma^\alpha_{\beta\gamma}$.  
(For simplicity the parallel transport is rendered as if the space is flat).
}
\end{figure}

Curvature of a Riemann manifold will cause a distortion in a vector field, say a coordinate field $X^\alpha(x)$, and this is depicted in Figure~\ref{parallel_transport}.  
The change in the coordinate field from one point $x$ to an adjacent point $x+\delta x$ is
\begin{equation}
\label{X_at_j}
X^\alpha (x+\delta x) = X^\alpha (x) + \underbrace{\delta x^\beta \partial_\beta X^\alpha }_{\delta X^\alpha(x)}.
\end{equation}
So, the change of the coordinate field due to the manifold is defined as
\begin{equation}
\delta X^\alpha(x) \equiv \delta x^\beta(x)\partial_\beta X^\alpha = X^\alpha(x+\delta x) - X^\alpha (x).
\end{equation}
The difference of the two coordinate vectors at point $j$ is 
\begin{equation}
[X^\alpha + \delta X^\alpha] - [X^\alpha + {\color{blue}\bar \delta X^\alpha}] = \delta X^\alpha(x) -{\color{blue}\bar \delta X^\alpha}(x).
\end{equation}
${\color{blue}\bar\delta X^\alpha}$ must vanish if either $\delta x^\alpha$ vanishes or $X^\alpha$ vanishes.
Therefore, we choose 
\begin{equation}
\label{parallel_dX}
{\color{blue}\bar\delta X^\alpha}= -\Gamma^\alpha_{\beta \gamma}(x) X^\beta(x)\delta x^\gamma,
\end{equation}
where $\Gamma^\alpha_{\beta \gamma}$ is a multiplicative factor, called the affine connection.  Its properties are yet to be determined.  At this stage, we understand it as a way to account for the curvature of the manifold.
%

The covariant derivative may be constructed as follows:
\begin{equation}
\label{covariant_derivative_definition}
\nabla_\gamma X^\alpha(x) \equiv
\frac{1}{\delta x^\gamma} 
\{ X^\alpha(x+\delta x) - [X^\alpha(x) + {\color{blue} \bar \delta X^\alpha}(x)]
\}.
\end{equation}
I do not use a limit in the definition to define the derivative.  Instead, I would like to just consider the situation where $\delta x^\gamma$ is a small finite quantity.  We will see below that this quantity drops out, justifying the form of (\ref{covariant_derivative_definition}).
Inserting (\ref{X_at_j}) and (\ref{parallel_dX}) into (\ref{covariant_derivative_definition}), yields
\begin{subequations}
\begin{eqnarray}
\nabla_\gamma X^\alpha(x) & = & \frac{1}{\delta x^\gamma} 
\{  X^\alpha (x) + {\delta x^\gamma \partial_\gamma X^\alpha } \\
\nonumber
&&- X^\alpha(x) +\Gamma^\alpha_{\beta \gamma}(x) X^\beta(x)\delta x^\gamma
\} \\
& = & \partial_\gamma X^\alpha(x) +\Gamma^\alpha_{\beta \gamma}(x) X^\beta(x).
\end{eqnarray}
\end{subequations}
So we see that $\delta x^\gamma$ cancels out and no limiting process to an infinitesimal size was really needed.  Dropping the explicit dependence on $x$, as this is to be understood,
we have the simple expression for the covariant derivative
\begin{equation}
\label{covariant_derivative}
\nabla_\gamma X^\alpha =  \partial_\gamma X^\alpha +\Gamma^\alpha_{\beta \gamma} X^\beta.
\end{equation}
In coordinate-based differential geometry, the covariant derivative of a tensor is given by its partial derivative plus correction terms, one for each index, involving an affine connection contracted with the tensor.  

\subsection{Spin connection}

In non-coordinate-based differential geometry, the ordinary affine connection coefficients $\Gamma^\lambda_{\mu \nu}$ are replaced by spin connection coefficients, denoted ${{\omega_\mu}^a}_b$, but otherwise the principle is the same. 
Each Latin index gets a correction factor that is the spin connection contracted with the tensor, for example
\begin{equation}
\nabla_\mu {X^a}_b
=
\partial_\mu {X^a}_b
+
{{\omega_\mu}^a}_c {X^c}_b
-
{{\omega_\mu}^c}_b {X^a}_c .
\end{equation}
The correction is positive for a upper index and negative for a lower index.
The spin connection is used to take covariant derivatives of spinors, whence its name.

The covariant derivative of a vector $X$ in the coordinate basis is
\begin{subequations}
\begin{eqnarray}
\nabla X
& = &
\left(
\nabla_\mu X^\nu
\right)
dx^\mu \otimes \partial_\nu \\
& = &
\label{affine_expansion}
\left(
\partial_\mu X^\nu
+
\Gamma^\nu_{\mu \lambda} X^\lambda
\right)
dx^\mu \otimes \partial_\nu.
\end{eqnarray}
\end{subequations}
The same object in a mixed basis, converted to the coordinate basis, is
%
\begin{subequations}
\begin{eqnarray}
\!\!\!
\nabla X
\!\!\!
& = &
\!\!\!
\left(
\nabla_\mu X^a
\right)
dx^\mu \otimes \hat{\bf e}_{(a)} \\
& = &
\left(
\partial_\mu X^a
+
{{\omega_\mu}^{a}}_b X^b
\right) dx^\mu \otimes \hat{\bf e}_{(a)}
\\
& = &
\left(
\partial_\mu
\left(
{e_\nu}^a X^\nu
\right)
+
{{\omega_\mu}^{a}}_b {e_\lambda}^b X^\lambda
\right)
dx^\mu \otimes 
\left(
{e^\sigma}_a \partial_\sigma
\right)
\\
\nonumber
& = &
{e^\sigma}_a
\left(
{e_\nu}^a \partial_\mu X^\nu
+
X^\nu \partial_\mu {e_\nu}^a
+
{{\omega_\mu}^{a}}_b {e_\lambda}^b X^\lambda
\right)
dx^\mu \otimes \partial_\sigma
\\
\\
\nonumber
& = &
\left(
\partial_\mu X^\sigma
+
{e^\sigma}_a
\partial_\mu {e_\nu}^a X^\nu
+
{e^\sigma}_a {e_\lambda}^b {{\omega_\mu}^{a}}_b X^\lambda
\right)
dx^\mu \otimes \partial_\sigma.
\\
\end{eqnarray}
\end{subequations}
Now, relabeling indices $\sigma \rightarrow \nu \rightarrow \lambda$ gives
%
\begin{eqnarray}
\nabla X
\nonumber
\!\!\!
& = &
\!\!\!
\left(
\partial_\mu X^\nu
+
{e^\nu}_a \partial_\mu {e_\lambda}^a X^\lambda
+
{e^\nu}_a {e_\lambda}^b {{\omega_\mu}^{a}}_b X^\lambda
\right)
dx^\mu \otimes \partial_\nu
\\
\nonumber
& = &
\label{affine_expansion_terms}
\left[
\partial_\mu X^\nu
+
\left(
{e^\nu}_a \partial_\mu {e_\lambda}^a
+
{e^\nu}_a {e_\lambda}^b {{\omega_\mu}^{a}}_b
\right) X^\lambda
\right]
dx^\mu \otimes \partial_\nu.\\
\end{eqnarray}
%
Therefore, comparing (\ref{affine_expansion}) with (\ref{affine_expansion_terms}), the affine connection in terms of the spin connection is
\begin{equation}
\Gamma^\nu_{\mu \lambda}
=
{e^\nu}_a \partial_\mu {e_\lambda}^a
+
{e^\nu}_a {e_\lambda}^b {{\omega_\mu}^{a}}_b.
\end{equation}
This can be solved for the spin connection
\begin{equation}
\label{spin_to_affine_connection_relation}
{{\omega_\mu}^{a}}_b
=
{e_\nu}^a {e^\lambda}_b
\Gamma^\nu_{\mu \lambda}
-
{e^\lambda}_b \partial_\mu {e_\lambda}^a.
\end{equation}

\subsection{Tetrad postulate}
\label{section:tetrad_postule}

The tetrad postulate is that the covariant derivative of the vierbein field vanishes, $\nabla_\mu {e_\nu}^a=0$, and this is merely a restatement of the relation we just found between the affine and spin connections (\ref{spin_to_affine_connection_relation}).
Left multiplying by $ {e_\nu}^b$ gives
\begin{subequations}
\begin{eqnarray}
{{\omega_\mu}^a}_b {e_\nu}^b
& = &
{e_\sigma}^a {e^\lambda}_b {e_\nu}^b
\Gamma^\sigma_{\mu \lambda}
-
{e^\lambda}_b {e_\nu}^b \partial_\mu {e_\lambda}^a
\\
& = &
{e_\sigma}^a \Gamma^\sigma_{\mu \nu}
-
\partial_\mu {e_\nu}^a.
\end{eqnarray}
\end{subequations}
Rearranging terms, we have the tetrad postulate
\begin{equation}
\label{tetrad_postulate}
\nabla_\mu {e_\nu}^a
\equiv
\partial_\mu {e_\nu}^a
-
{e_\sigma}^a \Gamma^\sigma_{\mu \nu}
+
{{\omega_\mu}^a}_b {e_\nu}^b
=
0.
\end{equation}

Let us restate (as a reminder) the correction rules for applying connections.  The covariant derivatives of a coordinate vector and 1-form are
\begin{subequations}
\begin{eqnarray}
\nabla_\mu X^\nu
& = &
\partial_\mu X^\nu
+
\Gamma^\nu_{\mu \lambda} X^\lambda
 \\
\nabla_\mu X_\nu
& = &
\partial_\mu X_\nu
-
\Gamma^\lambda_{\mu \nu} X_\lambda,
\end{eqnarray}
\end{subequations}
and similarly the covariant derivatives of a non-coordinate vector and 1-form are
\begin{subequations}
\begin{eqnarray}
\label{covariant_derivative_of_non_coordinate_1_form}
\nabla_\mu X^a
& = &
\partial_\mu X^a
+
{{\omega_\mu}^a}_b X^b\\
\nabla_\mu X_a
& = &
\partial_\mu X_a
-
{{\omega_\mu}^b}_a X_b.
\end{eqnarray}
\end{subequations}
We require a covariant derivative such as (\ref{covariant_derivative_of_non_coordinate_1_form}) to be Lorentz invariant
\begin{subequations}
\begin{eqnarray}
{\Lambda^{a'}}_a :  \nabla_\mu X^a
& \rightarrow &
\nabla_\mu 
\left(
{\Lambda^{a'}}_a X^a
\right)
\\
& = &
\left(
\nabla_\mu {\Lambda^{a'}}_a
\right)
X^a
+
{\Lambda^{a'}}_a \nabla_\mu X^a.
\end{eqnarray}
\end{subequations}
Therefore, the covariant derivative is Lorentz invariant,
\begin{eqnarray}
 \nabla_\mu X^a
 & = &
{\Lambda^{a'}}_a \nabla_\mu X^a,
\end{eqnarray}
so long as the covariant derivative of the Lorentz transformation vanishes,
\begin{equation}
\nabla_\mu {\Lambda^{a'}}_a = 0.
\end{equation}
This imposes a constraint that allows us to see how the spin connection behaves under a Lorentz transformation
\begin{equation}
\nabla_\mu {\Lambda^{a'}}_b
=
\partial_\mu {\Lambda^{a'}}_b
+
{{\omega_\mu}^{a'}}_c {\Lambda^c}_b
-
{{\omega_\mu}^{c}}_b {\Lambda^{a'}}_c
=
0,
\end{equation}
which we write as follows
\begin{equation}
{\Lambda^b}_{b'} \partial_\mu {\Lambda^{a'}}_b
+
{{\omega_\mu}^{a'}}_c {\Lambda^b}_{b'} {\Lambda^c}_b
-
{{\omega_\mu}^{c}}_b {\Lambda^b}_{b'} {\Lambda^{a'}}_c
=
0.
\end{equation}
Now ${\Lambda^b}_{b'} {\Lambda^c}_b = \delta^c_{b'}$, so we arrive at the transformation of the spin connection induced by a Lorentz transformation
\begin{equation}
{{\omega_\mu}^{a'}}_{b'}
=
{{\omega_\mu}^{c}}_b
{\Lambda^b}_{b'} {\Lambda^{a'}}_c
-
{\Lambda^b}_{b'} \partial_\mu {\Lambda^{a'}}_b.
\end{equation}
This means that the spin connection transforms inhomogeneously so that $\nabla_\mu X^a$ can transform like a Lorentz 4-vector.

The exterior derivative is defined as follows
\begin{subequations}
\begin{eqnarray}
{(dX)_{\mu \nu}}^a
& \equiv &
\nabla_\mu {X_\nu}^a
-
\nabla_\nu {X_\mu}^a
\\
\nonumber
& = &
\partial_\mu {X_\nu}^a
+
{{\omega_\mu}^a}_b {X_\nu}^b
-
\Gamma^\lambda_{\mu \nu} {X_\lambda}^a
\\
& -&
\partial_\nu {X_\mu}^a
-
{{\omega_\nu}^a}_b {X_\mu}^b
+ \Gamma^\lambda_{\nu \mu} {X_\lambda}^a\\
\nonumber
& = &
\partial_\mu {X_\nu}^a
-
\partial_\nu {X_\mu}^a
+
{{\omega_\mu}^a}_b {X_\nu}^b
-
{{\omega_\nu}^a}_b {X_\mu}^b.\\
\end{eqnarray}
\end{subequations}
Now, to make a remark about Cartan's notation as written in (\ref{covariant_othronormal_basis_coordinate_vector_component_2}),  one often writes the non-coordinate basis 1-form (\ref{noncoordinate_basis_1_form}) as
\begin{equation}
e^a
\equiv
\hat{\bf e}^{(a)}
=
{e_\mu}^a dx^\mu.
\end{equation}
The spin connection 1-form is 
\begin{equation}
{\omega^a}_b
=
{{\omega_\mu}^a}_b
dx^\mu.
\end{equation}
It is conventional to define a differential form
\begin{equation}
dA \equiv \partial_\mu A_\nu -\partial_\nu A_\mu
\end{equation}
and a wedge product
\begin{equation}
A\wedge B \equiv  A_\mu B_\nu -A_\nu B_\mu,
\end{equation}
which are both anti-symmetric in the Greek indices.
With this convention, originally due to \'Elie Cartan, the torsion can be written concisely in terms of the frame and spin connection 1-forms  as
\begin{equation}
T^a
=
de^a
+
{\omega^a}_b \wedge e^b.
\end{equation}
The notation is so compact that it is easy to misunderstand what it represents.  For example, writing the torsion explicitly in the coordinate basis we have
\begin{subequations}
\begin{eqnarray}
{T_{\mu \nu}}^\lambda
& = &
{e^\lambda}_a
{T_{\mu \nu}}^a
\\
\nonumber
& = &
{e^\lambda}_a
\left(
\partial_\mu {e_\nu}^a
-
\partial_\nu {e_\mu}^a
+
{{\omega_\mu}^a}_b {e_\nu}^b
-
{{\omega_\nu}^a}_b {e_\mu}^b
\right),
\\
& &
\end{eqnarray}
\end{subequations}
which fully expanded gives us
\begin{equation}
{T_{\mu \nu}}^\lambda
=
{e^\lambda}_a \partial_\mu {e_\nu}^a
+
{e^\lambda}_a {e_\nu}^b {{\omega_\mu}^a}_b
-
{e^\lambda}_a \partial_\nu {e_\mu}^a
-
{e^\lambda}_a {e_\mu}^b {{\omega_\nu}^a}_b.
\end{equation}
Since the affine connection is
\begin{equation}
\Gamma^\lambda_{\mu \nu}
=
{e^\lambda}_a \partial_\mu {e_\nu}^a
+
{e^\lambda}_a {e_\nu}^b {{\omega_\mu}^a}_b,
\end{equation}
the torsion then reduces to the simple expression
\begin{equation}
{T_{\mu \nu}}^\lambda
=
\Gamma^\lambda_{\mu \nu}
-
\Gamma^\lambda_{\nu \mu}.
\end{equation}
So, the torsion vanishes when the affine connection is symmetric in its lower two indices.

\section{Curvature}

We now derive the Riemann curvature tensor, and we do so in two ways.  The first way gives us an expression for the curvature in terms of the affine connection and the second way gives us an equivalent expression in terms of the spin connection.  The structure of both expressions are the same, so effectively the affine and spin connections can be interchanged, as long as one properly accounts for Latin and Greek indices.

\subsection{Riemann curvature from the affine connection}

\begin{figure*}[htbp]
\begin{center}
\setlength{\unitlength}{.5mm}
\begin{picture}(65,80)

\put(0,0){\line(1,5){10}}
\put(50,0){\line(1,5){10}}
\put(0,0){\line(1,0){50}}
\put(10,50){\line(1,0){50}}

\put(-8,-8){$x^\alpha$}
\put(53,-8){$x^\alpha+\delta x^\alpha$}
\put(53,55){$x^\alpha+\delta x^\alpha+dx^\alpha$}
\put(-8,55){$x^\alpha+dx^\alpha$}

\put(25,3){$\delta x^\alpha$}
\put(8,25){$dx^\alpha$}

\put(60,50){\circle*{2}}
\put(50,0){\circle*{2}}
\put(10,50){\circle*{2}}
\put(0,0){\circle*{2}}

\end{picture}
\vspace{0.2in}
\end{center}
\caption{\label{Plaquete}\footnotesize General plaquette of cell sizes $\delta x^\alpha$  and $dx^\alpha$ with its initial spacetime point at 4-vector $x_\alpha$ (bottom left corner).  The plaquette is a piece of a curved manifold.}
\end{figure*}
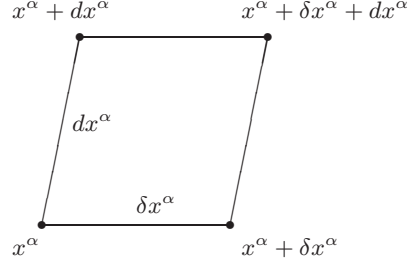

In this section, we will derive the Riemann curvature tensor in terms of the affine connection.  The development in this section follows the conventional approach of considering parallel transport around a  plaquette, as shown in Figure~\ref{Plaquete}.  (The term {\it plaquette} is borrowed from condensed matter theory and refers to a cell of a lattice.)  So, this is our first pass at understanding the origin of the Riemann curvature tensor.  In the following section, we will then re-derive the curvature tensor directly from the spin connection.

For the counterclockwise path ($x^\alpha \rightarrow x^\alpha +\delta x^\alpha \rightarrow x^\alpha +\delta x^\alpha +dx^\alpha$), we have:
\begin{subequations}
\begin{eqnarray}
X^\alpha(x+\delta x) &=&
X^\alpha(x) + {\color{blue}\bar \delta X^\alpha}(x)\\
\label{4_vector_connection}
&\stackrel{(\ref{parallel_dX})}{=}& X^\alpha(x) - \Gamma^\alpha_{\beta\gamma}(x) X^\beta(x) \delta x^\gamma. 
\end{eqnarray}
\end{subequations}
At the end point $x+\delta x + dx$, we have
\begin{subequations}
\begin{eqnarray}
\nonumber
X^\alpha(x+\delta x + dx) & = & 
X^\alpha(x+\delta x) 
+ {\color{blue}\bar \delta X^\alpha}(x+\delta x)\\
&&\\
\label{X_at_x_deltax_dx}
\nonumber
& \stackrel{(\ref{4_vector_connection})}{=} & X^\alpha(x) - \Gamma^\alpha_{\beta\gamma}(x) X^\beta(x)\delta x^\gamma\\
\nonumber
& &+ {\color{blue}\bar \delta X^\alpha}(x+\delta x).\\ 
\end{eqnarray}
\end{subequations}
Now, we need to evaluate the last term (parallel transport term) on the R.H.S.
%
\begin{widetext}
\begin{subequations}
\begin{eqnarray}
{\color{blue}\bar \delta X^\alpha}(x+\delta x) 
& \stackrel{(\ref{parallel_dX})}{=} & 
-\Gamma^\alpha_{\beta\gamma}(x+\delta x) X^\beta(x+\delta x) dx^\gamma\\
& \stackrel{(\ref{4_vector_connection})}{=} & -\left[\Gamma^\alpha_{\beta\gamma}(x)
 +\partial_\delta\Gamma^\alpha_{\beta\gamma}(x)
 \delta x^\delta\right]   \left[X^\beta(x) - \Gamma^\beta_{\mu\nu}(x) X^\mu(x) \delta x^\nu\right] dx^\gamma\\
 \label{parallel_transport_at_deltax_dx}
 \nonumber
 &=& -\Gamma^\alpha_{\beta\gamma} X^\beta dx^\gamma
  -\partial_\delta\Gamma^\alpha_{\beta\gamma} X^\beta \delta x^\delta dx^\gamma  
  +\Gamma^\alpha_{\beta\gamma}\Gamma^\beta_{\mu\nu} X^\mu \delta x^\nu dx^\gamma
 +
  \underbrace{ \partial_\delta\Gamma^\alpha_{\beta\gamma}\Gamma^\beta_{\mu\nu} X^\mu 
\delta x^\delta \delta x^\nu dx^\gamma}_{\text{neglect } 3^\text{rd} \text{ order term}},\\
\end{eqnarray}
\end{subequations}
where for brevity in the last expression we drop the explicit functional dependence on $x$, as this is understood.
Inserting (\ref{parallel_transport_at_deltax_dx}) into (\ref{X_at_x_deltax_dx}), the 4-vector at the end point is
\begin{equation}
X^\alpha(x+\delta x + dx)=
X^\alpha 
 -\Gamma^\alpha_{\beta\gamma}X^\beta \delta x^\gamma 
-\Gamma^\alpha_{\beta\gamma}X^\beta dx^\gamma
- \partial_\delta\Gamma^\alpha_{\beta\gamma} X^\beta \delta x^\delta dx^\gamma
+\Gamma^\alpha_{\beta\gamma}\Gamma^\beta_{\mu\nu} X^\mu \delta x^\nu dx^\gamma.
\end{equation}
Interchanging the indices of $\delta x$ and $dx$ in the last two terms, we have
\begin{equation}
X^\alpha(x+\delta x + dx) =
X^\alpha-\Gamma^\alpha_{\beta\gamma} X^\beta \delta x^\gamma
-\Gamma^\alpha_{\beta\gamma}X^\beta dx^\gamma 
-\partial_\gamma\Gamma^\alpha_{\beta\delta} X^\beta \delta x^\gamma dx^\delta 
+\Gamma^\alpha_{\beta\nu}\Gamma^\beta_{\mu\gamma} X^\mu \delta x^\gamma dx^\nu.
\end{equation}
Then, replacing $\nu$ with $\delta$ in this last term, we have
\begin{eqnarray}
\label{ccw_path}
X^\alpha(x+\delta x + dx) =
X^\alpha-\Gamma^\alpha_{\beta\gamma} X^\beta \delta x^\gamma 
-\Gamma^\alpha_{\beta\gamma}X^\beta dx^\gamma 
-\partial_\gamma\Gamma^\alpha_{\beta\delta} X^\beta \delta x^\gamma dx^\delta 
+\Gamma^\alpha_{\beta\delta}\Gamma^\beta_{\mu\gamma} X^\mu \delta x^\gamma dx^\delta.
\end{eqnarray}
For the clockwise path ($x^\alpha \rightarrow x^\alpha +dx^\alpha \rightarrow x^\alpha +dx^\alpha +\delta x^\alpha$), we get the same result as before with $dx^\alpha$ and $\delta x^\alpha$ interchanged:
\begin{eqnarray}
X^\alpha(x+\delta x + dx) =
X^\alpha-\Gamma^\alpha_{\beta\gamma} X^\beta dx^\gamma 
-\Gamma^\alpha_{\beta\gamma}X^\beta \delta x^\gamma 
-\partial_\gamma\Gamma^\alpha_{\beta\delta} X^\beta dx^\gamma \delta x^\delta 
+\Gamma^\alpha_{\beta\delta}\Gamma^\beta_{\mu\gamma} X^\mu dx^\gamma \delta x^\delta.
\end{eqnarray}
Interchanging the indices $\delta$ and $\gamma$ everywhere, we have
\begin{eqnarray}
\label{cw_path}
X^\alpha(x+\delta x + dx) & = & 
X^\alpha
-\Gamma^\alpha_{\beta \delta} X^\beta dx^\delta 
-\Gamma^\alpha_{\beta \delta}X^\beta \delta x^\delta 
-\partial_\delta\Gamma^\alpha_{\beta \gamma} X^\beta dx^\delta \delta x^\gamma
+\Gamma^\alpha_{\beta \gamma}\Gamma^\beta_{\mu\delta} X^\mu dx^\delta \delta x^\gamma,
\end{eqnarray}
which now looks like (\ref{ccw_path}) in the indices of the differentials.
Hence,  as we compute  (\ref{ccw_path}) minus (\ref{cw_path}), the zeroth and first order terms cancel, and the remaining second order terms  in (\ref{ccw_path}) and (\ref{cw_path}) add with the common factor $\delta x^\gamma dx^\delta$ (differential area)
\begin{subequations}
\begin{eqnarray}
\triangle X^\alpha &=& 
X^\alpha(x+\delta x + dx) - X^\alpha(x+dx+\delta x)\\
&=& 
\left(
\partial_\delta\Gamma^\alpha_{\beta\gamma} X^\beta
 - \partial_\gamma\Gamma^\alpha_{\beta\delta} X^\beta
 +\Gamma^\alpha_{\beta\delta}\Gamma^\beta_{\mu\gamma} X^\mu 
 -\Gamma^\alpha_{\beta\gamma}\Gamma^\beta_{\mu\delta} X^\mu
\right) 
\delta x^\gamma dx^\delta \\
&=& 
\left(
\partial_\delta\Gamma^\alpha_{\beta\gamma} 
 - \partial_\gamma\Gamma^\alpha_{\beta\delta} 
+\Gamma^\alpha_{\mu\delta}\Gamma^\mu_{\beta\gamma} 
-\Gamma^\alpha_{\mu\gamma}\Gamma^\mu_{\beta\delta} 
\right) 
X^\beta \delta x^\gamma dx^\delta.
\end{eqnarray}
\end{subequations}
\end{widetext}
%
So, the difference of transporting  the vector $X^\alpha$ along the two separate routes around the plaquette is related to the curvature of the manifold as follows
\begin{equation}
\triangle X^\alpha ={R^\alpha}_{\beta\delta\gamma} X^\beta \delta x^\gamma dx^\delta,
\end{equation}
and from here we arrive at our desired result and identify the Riemann curvature tensor as
\begin{equation}
\label{Riemann_curvature_tensor}
{R^\alpha}_{\beta\delta\gamma}\equiv
\partial_\delta\Gamma^\alpha_{\beta\gamma} 
 - \partial_\gamma\Gamma^\alpha_{\beta\delta} 
+\Gamma^\alpha_{\mu\delta}\Gamma^\mu_{\beta\gamma} 
-\Gamma^\alpha_{\mu\gamma}\Gamma^\mu_{\beta\delta}.
\end{equation}
Notice, from the identity (\ref{Riemann_curvature_tensor}), the curvature tensor is anti-symmetric in its last two indices, ${R^\alpha}_{\beta\delta\gamma}=-{R^\alpha}_{\beta\gamma\delta}$.

\subsection{Riemann curvature from the spin connection}

In this section, we will show that the Riemann curvature tensor (\ref{Riemann_curvature_tensor}) can be simply expressed in terms of the spin connection as follows:
\begin{equation}
{R^a}_b
=
d{\omega^a}_b
+
{\omega^a}_c \wedge {\omega^c}_b.
\end{equation}
Here the Greek indices are suppressed for brevity.   So, the first step is to explicitly write out the curvature tensor in all its indices and then to use the vierbein field to  convert the Latin indices to Greek indices, which gives us
\begin{equation}
\label{Riemann_curvature_tensor_spin_connection_form}
{R^\lambda}_{\sigma \mu \nu}
\equiv
{e^\lambda}_a {e_\sigma}^b
\left(
\partial_\mu {{\omega_\nu}^a}_b
-
\partial_\nu {{\omega_\mu}^a}_b
+
{{\omega_\mu}^a}_c {{\omega_\nu}^c}_b
-
{{\omega_\nu}^a}_c {{\omega_\mu}^c}_b
\right).
\end{equation}
The quantity in parentheses is a spin curvature.
Next, we will use (\ref{spin_to_affine_connection_relation}), which I restate here for convenience
\begin{equation}
\label{spin_to_affine_connection_relation_copy1}
{{\omega_\mu}^a}_b
=
{e_\rho}^a {e^\tau}_b
\Gamma^\rho_{\mu \tau}
-
{e^\tau}_b \partial_\mu {e_\tau}^a.
\end{equation}
Inserting (\ref{spin_to_affine_connection_relation_copy1}) into (\ref{Riemann_curvature_tensor_spin_connection_form}) gives
%
\begin{widetext}
\begin{eqnarray}
\label{curvature_expansion}
{R^\lambda}_{\sigma \mu \nu}
& = &
{e^\lambda}_a {e_\sigma}^b
\left[
\partial_\mu
\middle(
{e_\rho}^a {e^\tau}_b
\Gamma^\rho_{\nu \tau}
\middle)
-
\partial_\nu
\middle(
{e_\rho}^a {e^\tau}_b \Gamma^\rho_{\mu \tau}
\middle)
-
\partial_\mu {e^\tau}_b \partial_\nu {e_\tau}^a
+
\partial_\nu {e^\tau}_b \partial_\mu {e_\tau}^a
\right.
\\
\nonumber
& + &
\left(
{e_\rho}^a {e^\tau}_c \Gamma^\rho_{\mu \tau}
-
{e^\tau}_c \partial_\mu {e_\tau}^a
\middle)
\middle(
{e_{\rho'}}^c {e^{\tau'}}_b \Gamma^{\rho'}_{\nu \tau'}
-
{e^{\tau'}}_b \partial_\nu {e_{\tau'}}^c
\right)
\\
\nonumber
& - &
\left.
\left(
{e_\rho}^a {e^\tau}_c \Gamma^\rho_{\nu \tau}
-
{e^\tau}_c \partial_\nu {e_\tau}^a
\middle)
\middle(
{e_{\rho'}}^c {e^{\tau'}}_b \Gamma^{\rho'}_{\mu \tau'}
-
{e^{\tau'}}_b \partial_\mu {e_{\tau'}}^c
\right)
\right].
\end{eqnarray}
Reducing this expression is complicated to do.  Since the first term in (\ref{spin_to_affine_connection_relation_copy1}) depends on the affine connection $\Gamma$ and the second term depends on $\partial_\mu$, we will reduce (\ref{curvature_expansion}) in two passes, first considering 
terms that involve derivatives of the vierbein field and then terms that do not.

So as a first pass toward reducing (\ref{curvature_expansion}), we will consider all terms with derivatives of vierbeins, and show that these vanish.  To begin with, the first order derivative terms that appear with $\partial_\mu$ acting on vierbein fields are the following:
\begin{subequations}
\begin{eqnarray}
{e^\lambda}_a {e_\sigma}^b 
\left[
\partial_\mu
\left(
{e_\rho}^a {e^\tau}_b
\right)
\Gamma^\rho_{\nu \tau}
\right.
& - &
{e^\tau}_c
\left(
\partial_\mu {e_\tau}^a
\right)
{e_\rho}^c {e^\tau}_b
\Gamma^\rho_{\nu \tau}
+ 
{e_\rho}^a {e^\tau}_c \Gamma^\rho_{\nu \tau}
{e^{\tau'}}_b \partial_\mu 
\left. {e_{\tau'}}^c
\right]
\\
\nonumber
& = &
\left[
{e^\lambda}_a {e_\sigma}^b
{e^\tau}_b \partial_\mu {e_\rho}^a
+
{e^\lambda}_a {e_\sigma}^b {e_\rho}^a \partial_\mu {e^\tau}_b
\right]
\Gamma^\rho_{\nu \tau}
- 
{e^\lambda}_a {e_\sigma}^b {e^\tau}_c {e_\rho}^c {e^{\tau'}}_b
\left(
\partial_\mu {e_\tau}^a
\right)
\Gamma^\rho_{\nu \tau'}
\\
& + &
{e^\lambda}_a {e_\sigma}^b
{e_\rho}^a {e^\tau}_c {e^{\tau'}}_b \partial_\mu {e_{\tau'}}^c
\Gamma^\rho_{\nu \tau}
\\
& = &
{e^\lambda}_a \delta^\tau_\sigma
\partial_\mu {e_\rho}^a \Gamma^\rho_{\nu \tau}
+
\delta^\lambda_\rho {e_\sigma}^b \partial_\mu {e^\tau}_b 
\Gamma^\rho_{\nu \tau}
-
\delta^\tau_\rho \delta_\sigma^{\tau'}
{e^\lambda}_a \partial_\mu {e_\tau}^a 
\Gamma^\rho_{\nu \tau'}
+
\delta^\lambda_\rho \delta^{\tau'}_\sigma
{e^\tau}_c \partial_\mu {e_{\tau'}}^c
\Gamma^\rho_{\nu \tau}
\\
& = &
{e^\lambda}_a \partial_\mu {e_\rho}^a
\Gamma^\rho_{\nu \sigma}
+
{e_\sigma}^b \partial_\mu {e^\tau}_b 
\Gamma^\lambda_{\nu \tau}
-
{e^\lambda}_a \partial_\mu {e_\rho}^a
\Gamma^\rho_{\nu \sigma}
+
{e^\tau}_b \partial_\mu {e_\sigma}^b
\Gamma^\lambda_{\nu \tau}
\\
& = &
\partial_\mu
\left(
{e_\sigma}^b {e^\tau}_b
\right)
\Gamma^\lambda_{\nu \tau}
\\
&=&
\partial_\mu
\left(
\delta^\tau_\sigma
\right)
\Gamma^\lambda_{\nu \tau}
\\
&=&
0.
\end{eqnarray}
\end{subequations}
%
Similarly, all the first order derivative terms that appear with $\partial_\nu$ vanish as well. So, all the first order derivative terms vanish in (\ref{curvature_expansion}).  Next, we consider all second order derivatives, both $\partial_\mu$ and $\partial_\nu$, acting on vierbein fields.  All the terms with both $\partial_\mu$ and $\partial_\nu$ are the following
%
\begin{subequations}
\begin{eqnarray}
\nonumber
{e^\lambda}_a {e_\sigma}^b {e^\tau}_c {e^{\tau'}}_b
\left(
\partial_\mu {e_\tau}^a
\right)
\partial_\nu {e_{\tau'}}^c
& - & 
{e^\lambda}_a {e_\sigma}^b {e^\tau}_c {e^{\tau'}}_b
\left(
\partial_\nu {e_\tau}^a
\right)
\left(
\partial_\mu {e_{\tau'}}^c
\right)
-
{e^\lambda}_a {e_\sigma}^b
(\partial_\mu {e^\tau}_b) 
\partial_\nu {e_\tau}^a
+
{e^\lambda}_a {e_\sigma}^b
(\partial_\nu {e^\tau}_b) 
\partial_\mu {e_\tau}^a
\\
\nonumber
& = &
{e^\lambda}_a {e^\tau}_c
\left(
\partial_\mu {e_\tau}^a
\right)
\partial_\nu {e_\sigma}^c
-
{e^\lambda}_a {e^\tau}_c
\left(
\partial_\nu {e_\tau}^a
\right)
\partial_\mu {e_\sigma}^c
-
\left(
\partial_\nu {e^\lambda}_a
\right)
\left(
\partial_\mu {e_\sigma}^b
\right)
{e^\tau}_b {e_\tau}^a
\\
& + &
\left(
\partial_\mu {e^\lambda}_a
\right)
\left(
\partial_\nu {e_\sigma}^b
\right)
{e^\tau}_b {e_\tau}^a
\\
& = &
{e^\lambda}_a {e^\tau}_c
\left[
\left(
\partial_\mu {e_\tau}^a
\right)
\partial_\nu {e_\sigma}^c
-
\left(
\partial_\nu {e_\tau}^a
\right)
\partial_\mu {e_\sigma}^c
\right]
-
\partial_\nu {e^\lambda}_a
\partial_\mu {e_\sigma}^a
+
\partial_\mu {e^\lambda}_a
\partial_\nu {e_\sigma}^a
\\
& = &
-{e^\lambda}_a 
 {e_\tau}^a
\left(
\partial_\mu {e^\tau}_c
\right)
\partial_\nu {e_\sigma}^c
+
{e^\lambda}_a 
{e_\tau}^a
\left(
\partial_\nu {e^\tau}_c
\right)
\partial_\mu {e_\sigma}^c
-
\partial_\nu {e^\lambda}_a
\partial_\mu {e_\sigma}^a
+
\partial_\mu {e^\lambda}_a
\partial_\nu {e_\sigma}^a
\quad\\
& = &
-\delta^\lambda_\tau
\left(
\partial_\mu {e^\tau}_c
\right)
\partial_\nu {e_\sigma}^c
+
\delta^\lambda_\tau
\left(
\partial_\nu {e^\tau}_c
\right)
\partial_\mu {e_\sigma}^c
-
\partial_\nu {e^\lambda}_a
\partial_\mu {e_\sigma}^a
+
\partial_\mu {e^\lambda}_a
\partial_\nu {e_\sigma}^a
\\
& = &
-
\left(
\partial_\mu {e^ \lambda}_c
\right)
\partial_\nu {e_\sigma}^c
+
\left(
\partial_\nu {e^\lambda}_c
\right)
\partial_\mu {e_\sigma}^c
-
\partial_\nu {e^\lambda}_a
\partial_\mu {e_\sigma}^a
+
\partial_\mu {e^\lambda}_a
\partial_\nu {e_\sigma}^a
\\
& = &
0.
\end{eqnarray}
\end{subequations}
%
Hence, all the second order derivative terms in (\ref{curvature_expansion}) vanish, as do the first order terms.  Note that we made use of the fact $\partial_\mu
\left(
{e^\lambda}_a {e_\tau}^a
\right)
=0$, so as to swap the order of differentiation,  
\begin{equation}
\label{vierbien_derivative_identity}
\partial_\mu
\left(
{e^\lambda}_a
\right) {e_\tau}^a
=-{e^\lambda}_a
\left(
\partial_\mu {e_\tau}^a
\right).
\end{equation}
Finally, as a second pass toward reducing (\ref{curvature_expansion}) to its final form, we now consider all the remaining terms (no derivatives of the vierbein fields), and these lead to the curvature tensor expressed solely as a function of the affine connection:
%
\begin{subequations}
\begin{eqnarray}
{R^\lambda}_{\sigma \mu \nu}
& = &
{e^\lambda}_a {e_\sigma}^b
\left[
{e_\rho}^a {e^\tau}_b
\left(
\partial_\mu \Gamma^\rho_{\nu \tau}
-
\partial_\nu \Gamma^\rho_{\mu \tau}
\right)
+
{e_\rho}^a {e^{\tau'}}_b
\left(
\Gamma^\rho_{\mu \tau} \Gamma^\tau_{\nu \tau'}
-
\Gamma^\rho_{\nu \tau} \Gamma^\tau_{\mu \tau'}
\right)
\right]
\\
& = &
\delta^\lambda_\rho \delta^\tau_\sigma
\left(
\partial_\mu \Gamma^\rho_{\nu \tau}
-
\partial_\nu \Gamma^\rho_{\mu \tau}
\right)
+
\delta_\rho^\lambda \delta^{\tau'}_\sigma
\left(
 \Gamma^\rho_{\mu \tau}  \Gamma^\tau_{\nu \tau'}
-
\Gamma^\rho_{\nu \tau} \Gamma^\tau_{\mu \tau'}
\right).
\end{eqnarray}
\end{subequations}
\end{widetext}
%
Applying the Kronecker deltas, we arrive at the final result 
\begin{equation}
\label{Riemann_curvature_tensor_again}
{R^\lambda}_{\sigma \mu \nu}
=
\partial_\mu
\Gamma^\lambda_{\nu \sigma}
-
\partial_\nu
\Gamma^\lambda_{\mu \sigma}
+
\Gamma^\lambda_{\mu \tau}
\Gamma^\tau_{\nu \sigma}
-
\Gamma^\lambda_{\nu \tau}
\Gamma^\tau_{\mu \sigma},
\end{equation}
which is identical to (\ref{Riemann_curvature_tensor}).  If we had not already derived the curvature tensor, 
we could have written  (\ref{Riemann_curvature_tensor_again})  down by inspection because of its similarity to (\ref{Riemann_curvature_tensor_spin_connection_form}), essentially replacing the spin connection with the affine connection.

\section{Mathematical constructs}


Here we assemble a number of preliminary identities that we will use later to derive the Einstein equation.  An identity we will need allows us to evaluate the trace of $M^{-1}\partial_\mu M$ where $M$ is a 2-rank tensor
\begin{equation}
\label{trace_matrix_derivative_identity}
Tr[ M^{-1}\partial_\mu M] = \partial_\mu \ln |M|.
\end{equation}
As an example of this identity, consider the following $2\times 2$ matrix and its inverse
\begin{equation}
M
=
\begin{pmatrix}
   a   &  b  \\
   c   &  d
\end{pmatrix}
\qquad
M^{-1}
=
\frac{1}{|M|}
\begin{pmatrix}
   d   &  -b  \\
   -c   &  a
\end{pmatrix}.
\end{equation}
A demonstration of the trace identity (\ref{trace_matrix_derivative_identity}) for the simplest case of one spatial dimension is
%
\begin{widetext}
\begin{subequations}
\label{trace_similarity_transform_of_derivative_identity}
\begin{eqnarray}
\text{Tr}
\left[
M^{-1} \partial_x M
\right]
& = &
\text{Tr}
\left[
\frac{1}{a d - b c}
\begin{pmatrix}
   d   &  -b  \\
   -c   &  a
\end{pmatrix}
\begin{pmatrix}
 \partial_x  a   & \partial_x b  \\
 \partial_x  c   & \partial_x d
\end{pmatrix}
\right]
\\
& = &
\text{Tr}
\left[
\frac{1}{a d - b c}
\begin{pmatrix}
 \partial_x  a \, d - b \partial_x c  & \partial_x b \, d - b \partial_x d  \\
 - \partial_x a \, c + a \partial_x  c  & - \partial_x b \, c + a \partial_x d
\end{pmatrix}
\right]
\\
& = &
\frac{1}{a d - b c}
\left[
\partial_x(a d) - \partial_x(b c)
\right]
\\
& = &
\frac{\partial_x(a d - b c)}{a d - b c}
\\
& = &
\partial_x \ln |M|.
\end{eqnarray}
\end{subequations}
\end{widetext}
%
This identity holds for matrices of arbitrary size.  In our case, we shall need this identity for the case of $4\times 4$ matrices.

The contravariant and covariant metric tensors are orthogonal
\begin{equation}
g^{\lambda \mu} g_{\mu \nu}
=
\delta^\lambda_\nu,
\end{equation}
so
\begin{equation}
g^{\mu\nu} \rightarrow (g_{\mu \nu})^{-1}.
\end{equation}
We also define the negative determinant of the metric tensor
\begin{equation}
g \equiv - \text{Det} \, g_{\mu \nu}.
\end{equation}

\subsection{Consequence of tetrad postulate}

The tetrad postulate of Section~\ref{section:tetrad_postule} is that the vierbein field is invariant under parallel transport
\begin{equation}
\label{tetrad_postulate}
\nabla_\mu {e_\nu}^a = 0.
\end{equation}
That the metric tensor is invariant under parallel transport then immediately follows 
\begin{subequations}
\label{invariant_g_mu_nu_under_parallel_transport}
\begin{eqnarray}
\nabla_\mu g_{\nu \lambda}
& = &
\nabla_\mu
\left(
{e_\nu}^a {e_\lambda}^c n_{a b}
\right)
\\
& = &
\left(
\nabla_\mu {e_\nu}^a
\right)
{e_\lambda}^b n_{a b}
+
{e_\nu}^a
\left(
\nabla_\mu {e_\lambda}^b
\right)
n_{a b}
\qquad
\\
& \stackrel{(\ref{tetrad_postulate})}{=} &
0.
\end{eqnarray}
\end{subequations}
This is called metric compatibility.

\subsection{Affine connection in terms of the metric tensor}

Now we can make use of (\ref{invariant_g_mu_nu_under_parallel_transport}) to compute the affine connection.  Permuting indices, we can write
\begin{subequations}
\begin{equation}
\label{vanishing_covariant_derivative_metric_tensor_1}
\nabla_\rho g_{\mu \nu}
=
\partial_\rho g_{\mu\nu}
-
\Gamma^\lambda_{\rho \mu}
g_{\lambda \nu}
-
\Gamma^\lambda_{\rho \nu}
g_{\mu \lambda}
=
0
\end{equation}
\begin{equation}
\label{vanishing_covariant_derivative_metric_tensor_2}
\nabla_\mu g_{\nu \rho}
=
\partial_\mu g_{\nu \rho}
-
\Gamma^\lambda_{\mu \nu}
g_{\lambda \rho}
-
\Gamma^\lambda_{\mu \rho}
g_{\nu \lambda}
=
0
\end{equation}
\begin{equation}
\label{vanishing_covariant_derivative_metric_tensor_3}
\nabla_\nu g_{\rho \mu}
=
\partial_\nu g_{\rho \mu}
-
\Gamma^\lambda_{\nu \rho}
g_{\lambda \mu}
-
\Gamma^\lambda_{\nu \mu}
g_{\rho \lambda}
=
0.
\end{equation}
\end{subequations}
Now, we take (\ref{vanishing_covariant_derivative_metric_tensor_1}) $-$  (\ref{vanishing_covariant_derivative_metric_tensor_2}) $-$ (\ref{vanishing_covariant_derivative_metric_tensor_3}):
\begin{equation}
\partial_\rho g_{\mu\nu}
-
\partial_\mu g_{\nu \rho}
-
\partial_\nu g_{\rho \mu}
+
2 \Gamma^\lambda_{\mu \nu}
g_{\lambda \rho}
=
0.
\end{equation}
Multiplying through by $g^{\sigma \rho}$ allows us to solve for the affine connection
\begin{equation}
\Gamma^\sigma_{\mu \nu}
=
\frac{1}{2} g^{\sigma \rho}
\left(
\partial_\mu g_{\nu \rho}
+
\partial_\nu g_{\rho \mu}
-
\partial_\rho g_{\mu \nu}
\right).
\end{equation}
Then, contracting the $\sigma$ and $\mu$ indices, we have
\begin{subequations}
\begin{eqnarray}
\Gamma^\mu_{\mu \nu}
& = &
\frac{1}{2} g^{\mu \rho}
\left(
\partial_\mu g_{\nu \rho}
+
\partial_\nu g_{\rho \mu}
-
\partial_\rho g_{\mu \nu}
\right)
\\
& = &
\frac{1}{2} g^{\mu \rho}
\left(
\partial_\nu g_{\rho \mu}
+
\partial_\mu g_{\rho \nu}
-
\partial_\rho g_{\mu \nu}
\right)
\\
& = &
\frac{1}{2} g^{\mu \rho}
\partial_\nu g_{\rho \mu}+
\frac{1}{2} g^{\mu \rho}
\left\{
\partial_\mu g_{\rho \nu}
-
\partial_\rho g_{\mu \nu}
\right\}.
\end{eqnarray}
\end{subequations}
Since the metric tensor is symmetric and the last term in brackets is anti-symmetric in the $\mu\,\rho$ indices, the product must vanish.  Thus
\begin{equation}
\label{contracted_affine_connection_as_similarity_transformation_of_covariant_derivative}
\Gamma^\mu_{\mu \nu}
=
\frac{1}{2} g^{\mu \rho}
\partial_\nu g_{\rho \mu}.
\end{equation}
Furthermore, rewriting (\ref{contracted_affine_connection_as_similarity_transformation_of_covariant_derivative}) as the trace of the similarity transformation of  the covariant derivative $\text{Tr}[M^{-1} \partial_\nu M] = \partial_\nu \ln \text{Det} \, M$ that we demonstrated in (\ref{trace_similarity_transform_of_derivative_identity}), we have
\begin{subequations}
\label{once_contracted_affine_connection}
\begin{eqnarray}
\Gamma^\mu_{\mu \nu}
& = &
\frac{1}{2}
\text{Tr}
\left(
g^{\lambda \rho} \partial_\nu g_{\rho \mu}
\right)
\\
& = &
\frac{1}{2}
\partial_\nu \ln \text{Det} \, g_{\rho \mu}
\\
& = &
\frac{1}{2}
\partial_\nu \ln(- g),
\qquad
g \equiv - \text{Det} \, g_{\rho \mu}
\\
& = &
\partial_\nu \ln \sqrt{- g}
\\
\label{once_contracted_affine_connection_final}
& = &
\frac{1}{\sqrt{- g}}
\partial_\nu \sqrt{- g}.
\end{eqnarray}
\end{subequations}
Equating (\ref{contracted_affine_connection_as_similarity_transformation_of_covariant_derivative}) to (\ref{once_contracted_affine_connection_final}), we have
\begin{equation}
\partial_\nu \sqrt{- g}
=
\frac{1}{2} \sqrt{- g}\,
 g^{\mu \rho}
\partial_\nu g_{\rho \mu}.
\end{equation}
For generality, we write this corollary to (\ref{once_contracted_affine_connection_final}) as follows:
\begin{equation}
\label{variation_of_sqrt_of_minus_determinant}
\delta \sqrt{- g}
=
\frac{1}{2} \sqrt{- g} \,
g^{\mu \nu} \delta \, g_{\mu \nu}.
\end{equation}

\subsection{Invariant volume element}

Now, we consider the transport of a general 4-vector
\begin{equation}
\nabla_\nu V^\mu
=
\partial_\nu V^\mu
+
\Gamma^\mu_{\nu \lambda} V^\lambda.
\end{equation}
Therefore, the 4-divergence of $V^\mu$ is
\begin{subequations}
\begin{eqnarray}
\nabla_\mu V^\mu
& = &
\partial_\mu V^\mu
+
\Gamma^\mu_{\mu \lambda} V^\lambda
\\
& \stackrel{(\ref{once_contracted_affine_connection_final})}{=} &
\partial_\mu V^\mu
+
\frac{1}{\sqrt{- g}}
\left(
\partial_\lambda \sqrt{- g}
\right) V^\lambda
\\
\label{4_divergence_of_V_mu}
& = &
\frac{1}{\sqrt{- g}}
\partial_\mu
\left(
\sqrt{- g} V^\mu
\right).
\end{eqnarray}
\end{subequations}
If $V^\mu$ vanishes at infinity, then integrating over all space yields
\begin{equation}
\int d^4x \sqrt{- g} \, \nabla_\mu V^\mu
=
\int d^4x \, \partial_\mu
\left(
\sqrt{- g} V^\mu
\right)
=
0,
\end{equation}
which is a covariant form of Gauss's theorem where the invariant volume element is
\begin{equation}
dV = \sqrt{- g} \, d^4x.
\end{equation}

\subsection{Ricci tensor}

The Ricci tensor is the second rank tensor formed from the Riemann curvature tensor as follows:
\begin{equation}
R_{\sigma \nu}
\equiv
R^\lambda_{\sigma \lambda \nu}
=
g^{\lambda \mu} R_{\mu \sigma \lambda \nu},
\end{equation}
where the Riemann tensor is
\begin{equation}
{R^\rho}_{\sigma \mu \nu}
=
\partial_\mu \Gamma^\rho_{\nu \sigma}
-
\partial_\nu \Gamma^\rho_{\mu \sigma}
+
\Gamma^\rho_{\mu \lambda} \Gamma^\lambda_{\nu \sigma}
-
\Gamma^\rho_{\nu \lambda} \Gamma^\lambda_{\mu \sigma}.
\end{equation}
Therefore, the Ricci tensor can be written as
\begin{subequations}
\begin{eqnarray}
R_{\sigma \nu}
& = &
\partial_\rho \Gamma^\rho_{\nu \sigma}
-
\partial_\nu \Gamma^\rho_{\rho \sigma}
+
\Gamma^\rho_{\rho \lambda} \Gamma^\lambda_{\nu \sigma}
-
\Gamma^\rho_{\nu \lambda} \Gamma^\lambda_{\rho \sigma}
\\
\label{Ricci_tensor_form1}
& = &
\partial_\rho \Gamma^\rho_{\nu \sigma}
-
\Gamma^\lambda_{\nu \rho} \Gamma^\rho_{\lambda \sigma}
-
\partial_\nu \Gamma^\rho_{\sigma \rho}
+
\Gamma^\lambda_{\nu \sigma} \Gamma^\rho_{\lambda \rho}.
\qquad
\end{eqnarray}
\end{subequations}
Now, using the correction for a covariant vector
\begin{equation}
\nabla_\rho A_\nu 
=
\partial_\rho A_\nu
-
\Gamma^\lambda_{\rho \nu} A_\lambda,
\end{equation}
we can write (\ref{Ricci_tensor_form1}) as
\begin{equation}
\label{Palatini_identity}
R_{\sigma \nu}
=
\nabla_\rho \Gamma^\rho_{\nu \sigma}
-
\nabla_\nu \Gamma^\rho_{\sigma \rho}.
\end{equation}
(\ref{Palatini_identity}) is known as the Palatini identity.
The scalar curvature is the following contraction of the Ricci tensor
\begin{equation}
R \equiv g^{\mu \nu} R_{\mu \nu}.
\end{equation}

\section{Gravitational action}

\subsection{Free field gravitational action}

The action for the source free gravitational field is
\begin{equation}
\label{gravitational_action}
I_G
=
\frac{1}{16 \pi G}
\int d^4x \, \sqrt{- g}\, R(x).
\end{equation}
The equation of motion for the metric tensor can be determined by varying (\ref{gravitational_action}) with respect to the metric tensor field.  The variation is carried out in several stages.
The variation of the Lagrangian density is
\begin{equation}
\label{variation_of_metric}
\delta
\left(
\sqrt{- g} R
\right)
=
\sqrt{- g} R_{\mu \nu} \delta g^{\mu \nu}
+
R\, \delta \sqrt{- g}
+
\sqrt{- g} \, g^{\mu \nu} \delta R_{\mu \nu}.
\end{equation}
From the Palatini identity (\ref{Palatini_identity}), the change in the Ricci tensor can be written as
\begin{equation}
\label{variation_of_Ricci_tensor}
\delta R_{\mu \nu}
=
- \nabla_\nu \delta \Gamma^\lambda_{\mu \lambda}
+
\nabla_\lambda \delta \Gamma^\lambda_{\mu \nu},
\end{equation}
where we commute the variational change with the covariant derivative.
Now, we can expand the third term  on the R.H.S. of (\ref{variation_of_metric}) as follows:
%
\begin{widetext}
\begin{subequations}
\begin{eqnarray}
\sqrt{- g} \, g^{\mu \nu} \delta R_{\mu \nu}
\nonumber
& \stackrel{(\ref{variation_of_Ricci_tensor})}{=} &
- \sqrt{- g}
\left[
g^{\mu \nu} \nabla_\nu \delta \Gamma^\lambda_{\mu \lambda}
-
g^{\mu \nu} \nabla_\lambda \delta \Gamma^\lambda_{\mu \nu}
\right]
\\
& &
\\
\nonumber
& \stackrel{(\ref{invariant_g_mu_nu_under_parallel_transport})}{=} &
- \sqrt{- g}
\left[
\nabla_\nu
\underbrace{\left( g^{\mu \nu} \delta \Gamma^\lambda_{\mu \lambda} \right)}_{\text{like} \, V^\nu}
-
\nabla_\lambda
\underbrace{\left( g^{\mu \nu} \delta \Gamma^\lambda_{\mu \nu} \right)}_{\text{like} \, V^\mu}
\right],
\\ 
& &
\end{eqnarray}
%
since $\nabla_\mu g_{\nu \rho} = 0$.
Now from (\ref{4_divergence_of_V_mu}) we know $\nabla_\nu V^\nu = \frac{1}{\sqrt{- g}} \partial_\nu (\sqrt{- g} V^\nu)$, so for the variation of the third term we have
\begin{equation}
\sqrt{- g} \, g^{\mu \nu} \delta R_{\mu \nu}
=
- \partial_\nu
\left(
\sqrt{- g} \, g^{\mu \nu} \delta \Gamma^\lambda_{\mu \lambda}
\right)
+
\partial_\lambda
\left(
\sqrt{- g} \, g^{\mu \nu} \delta \Gamma^\lambda_{\mu \nu}
\right).
\end{equation}
\end{subequations}
These surface terms drop out when integrated over all space, so the third term on the R.H.S. of (\ref{variation_of_metric}) vanishes.  Finally, inserting the result (\ref{variation_of_sqrt_of_minus_determinant})
\begin{equation*}
\delta \sqrt{- g}
=
\frac{1}{2} \sqrt{- g} \, g^{\mu \nu} \delta g_{\mu \nu}
\end{equation*}
into the second term on the R.H.S. of (\ref{variation_of_metric})
we can write the variation of the gravitational action (\ref{gravitational_action}) entirely in terms of the variation of the metric tensor field
\begin{equation}
\delta I_G
=
\frac{1}{16 \pi G}
\int d^4x \, \sqrt{- g}
\left[
R_{\mu \nu} \delta g^{\mu \nu}
+
\frac{1}{2} g^{\mu \nu} R \,\delta g_{\mu \nu}
\right].
\end{equation}
The variation of identity vanishes,
\begin{equation}
\delta[\delta^\mu_\lambda]
=
\delta
\left[
g^{\mu \tau} g_{\tau \lambda}
\right]
=
\left(
\delta g^{\mu \tau}
\right)
g_{\tau \lambda}
+
g^{\mu \tau} \delta g_{\tau \lambda}
=
0,
\end{equation}
from which we find the following useful identity
\begin{equation}
\delta g^{\mu \tau} g^{\nu \lambda} g_{\tau \lambda}
+
g^{\nu \lambda} g^{\mu \tau} \delta g_{\tau \lambda}
=
0
\end{equation}
or
\begin{equation}
\delta g^{\mu \nu}
=
-  g^{\nu \lambda} g^{\mu \tau} \delta g_{\tau \lambda}.
\end{equation}
With this identity, we can write the variation of the free gravitational action as
%
\begin{subequations}
\begin{eqnarray}
\delta I_G
\nonumber
& = &
- \frac{1}{16 \pi G}
\int d^4x \, \sqrt{- g}
\left[
R_{\mu \nu} g^{\mu \tau} g^{\nu \lambda}  \delta g_{\tau \lambda}
-
\frac{1}{2} g^{\mu \nu} R \delta g_{\mu \nu}
\right]
\\
& &
\\
\label{variation_of_source_free_gravitational_action_wrt_g_mu_nu}
& = &
- \frac{1}{16 \pi G}
\int d^4x \, \sqrt{- g}
\left[
R^{\mu \nu}
-
\frac{1}{2} g^{\mu \nu} R
\right]
\delta g_{\mu \nu}.
\end{eqnarray}
\end{subequations}
%
Since the variation of the metric does not vanish in general, for the gravitational action to vanish the quantity in square brackets must vanish.  This quantity is call the Einstein tensor
\begin{equation}
G^{\mu \nu}
\equiv
R^{\mu \nu}
-
\frac{1}{2} g^{\mu \nu} R.
\end{equation}
So, the equation of motion for the free gravitation field is simply
\begin{equation}
\label{Einstein_equation_for_free_gravitational_field}
G^{\mu \nu}=0.
\end{equation}

\subsection{Variation with respect to the vierbein field}

In terms of the vierbein field, the metric tensor is
\begin{equation}
g_{\mu \nu}(x)
=
{e_\mu}^a(x) {e_\nu}^b(x) \eta_{a b},
\end{equation}
so its variation can be directly written in terms of the variation of the vierbein
\begin{subequations}
\begin{eqnarray}
\delta g_{\mu \nu}
& = &
\delta {e_\mu}^a {e_\nu}^b \eta_{a b}
+
{e_\mu}^a \delta {e_\nu}^b \eta_{a b}
\\
\label{variation_of_metric_in_terms_of_vierbein_field_positive}
& = &
\delta {e_\mu}^a e_{\nu a}
+
e_{\mu a} \delta {e_\nu}^a
\\
& \stackrel{(\ref{vierbien_derivative_identity})}{=} &
- {e_\mu}^a \delta e_{\nu a}
-
\delta e_{\mu a} {e_\nu}^a.
\end{eqnarray}
Now, we can look upon $\delta e_{\mu a}$ as a field quantity whose indices we can raise or lower with the appropriate use of the metric tensor.  Thus, we can write the variation of the metric as
\begin{eqnarray}
\delta g_{\mu \nu}
& = &
- g_{\nu \lambda} {e_\mu}^a \delta {e^\lambda}_a
-
g_{\mu \lambda} \delta {e^\lambda}_a {e_\nu}^a
\\
& = &
\label{variation_of_metric_in_terms_of_vierbein_field}
- \left(
g_{\mu \lambda} {e_\nu}^a
+
g_{\nu \lambda} {e_\mu}^a
\right)
\delta {e^\lambda}_a.
\end{eqnarray}
\end{subequations}
Therefore, inserting this into (\ref{variation_of_source_free_gravitational_action_wrt_g_mu_nu}), the variation of the source-free gravitational action with respect to the vierbein field is
%
\begin{subequations}
\begin{eqnarray}
\delta I_G
& = &
\frac{1}{16 \pi G}
\int d^4x \sqrt{- g}
\left(
R^{\mu \nu}
-
\frac{1}{2} g^{\mu \nu} R
\middle)
\middle(
g_{\mu \lambda} {e_\nu}^a
+
g_{\nu \lambda} {e_\mu}^a
\right)
\delta {e^\lambda}_a
\\
& = &
\frac{1}{16 \pi G}
\int d^4x \sqrt{- g}
\left(
{R_\lambda}^\nu {e_\nu}^a
-
\frac{1}{2} \delta^\nu_\lambda R \, {e_\nu}^a
+
{R^\mu}_\lambda {e_\mu}^a
-
\frac{1}{2} \delta^\mu_\lambda R \, {e_\mu}^a
\right)
\delta {e^\lambda}_a\\
&=&
\label{variation_of_source_free_gravitational_action}
\frac{1}{8 \pi G}
\int d^4x \sqrt{- g}
\left[
\left(
{R^\mu}_\lambda
-
\frac{1}{2} \delta^\mu_\lambda R
\right)
{e_\mu}^a
\right]
 \delta {e^\lambda}_a.
\end{eqnarray}
\end{subequations}
\end{widetext}
%
Since the variation of the vierbein field does not vanish in general, for the gravitational action to vanish the quantity in square brackets must vanish.  Multiplying this by $g^{\lambda\nu}$, the equation of motion is 
\begin{equation}
\label{Einstein_equation_for_free_gravitational_field_vierbein_representation}
G^{\mu\nu}{e_\mu}^a =0,
\end{equation}
which leads to (\ref{Einstein_equation_for_free_gravitational_field}) since ${e_\mu}^a\ne 0$.  Yet (\ref{Einstein_equation_for_free_gravitational_field_vierbein_representation}) is a more general equation of motion since it allows cancelation across components instead of the simplest case where each component of $G^{\mu\nu}$ vanishes separately.

\subsection{Action for a gravitational source}

The fundamental principle in general relativity is that the presence of matter warps the spacetime manifold in the vicinity of the source.    The vierbein field allows us to quantify this principle in a rather direct way.  The variation of the action for the matter source to lowest order is linearly proportional to the variation of the vierbein field
\begin{equation}
\label{fundamental_variation_of_the_matter_action}
\delta I_M
=
\int d^4x \sqrt{- g} \, {u_\lambda}^a \delta {e^\lambda}_a ,
\end{equation}
where the components ${u_\lambda}^a$ are constants of proportionality.  However, the usual definition of the matter action is as a functional derivative with respect to the metric
\begin{equation}
\label{matter_action_in_terms_of_energy_momentum_tensor}
\frac{\delta I_M}{\delta g_{\mu\nu}} \equiv \frac{1}{2}\int d^4 x \sqrt{-g} \,T^{\mu\nu}.
\end{equation}
So, in consideration of (\ref{fundamental_variation_of_the_matter_action}) and (\ref{matter_action_in_terms_of_energy_momentum_tensor}), we should write 
\begin{subequations}
\begin{eqnarray}
 {u_\lambda}^a \delta {e^\lambda}_a &=&  \frac{1}{2}\,T^{\mu\nu} \, \delta g_{\mu\nu} \\
 &\stackrel{(\ref{variation_of_metric_in_terms_of_vierbein_field})}{=}&
-  \frac{1}{2}\,T^{\mu\nu}
   \left(
g_{\mu \lambda} {e_\nu}^a
+
g_{\nu \lambda} {e_\mu}^a
\right)
\delta {e^\lambda}_a,
\qquad
  \end{eqnarray}
\end{subequations}
which we can solve for $T^{\mu\nu}$.  Dividing out $\delta {e^\lambda}_a$ and then multiplying through by $g^{\lambda\beta}$ we get
\begin{subequations}
\begin{eqnarray}
 {u}^{\beta a} &=&  
 -  \frac{1}{2}\,T^{\mu\nu}
   \left(
\delta^\beta_{\mu} {e_\nu}^a
+
\delta^\beta_{\nu} {e_\mu}^a
\right)\\
&=&
 -  \frac{1}{2}\
   \left(
T^{\beta\nu} {e_\nu}^a
+
T^{\mu\beta} {e_\mu}^a
\right)
\\
&=&
-T^{\beta\nu}{e_\nu}^a,
\end{eqnarray}
\end{subequations}
since the energy-momentum tensor is symmetric.  Thus, we have
\begin{equation}
T^{\mu\nu} = -  {u}^{\mu a} {e^\nu}_a.
\end{equation}
Alternatively, again in consideration of (\ref{fundamental_variation_of_the_matter_action}) and (\ref{matter_action_in_terms_of_energy_momentum_tensor}), we could also write 
\begin{subequations}
\begin{eqnarray}
 {u_\lambda}^a \delta {e^\lambda}_a &=&  \frac{1}{2}\,T^{\mu\nu} \, \delta g_{\mu\nu} \\
 &\stackrel{(\ref{variation_of_metric_in_terms_of_vierbein_field_positive})}{=}&
 \frac{1}{2}\,T^{\mu\nu}
   \left(\delta {e_\mu}^a e_{\nu a}
+
e_{\mu a} \delta {e_\nu}^a
\right)
\\
&=&
 \frac{1}{2}\,T_{\mu\nu}
   \left(\delta {e^\mu}_a e^{\nu a}
+
e^{\mu a} \delta {e^\nu}_a
\right)
\\
&=&
 \frac{1}{2}\,
   \left(T_{\lambda\nu}\delta {e^\lambda}_a e^{\nu a}
+
T_{\mu\lambda}
e^{\mu a} \delta {e^\lambda}_a
\right)
\qquad
\\
&=&
 \frac{1}{2}\,
   \left(T_{\lambda\nu} e^{\nu a}
+
T_{\mu\lambda}
e^{\mu a}
\right)\delta {e^\lambda}_a
\\
&=&
T_{\lambda\nu} e^{\nu a}
\delta {e^\lambda}_a.
  \end{eqnarray}
\end{subequations}
This implies that the energy-stress tensor is proportional to the vierbein field
\begin{equation}
T_{\mu \nu}
=
e_{\mu a} {u_\nu}^a.
\end{equation}
Consequently, the variation of the energy-stress tensor is then
\begin{equation}
\delta T_{\mu \nu}
=
\delta e_{\mu a} {u_\nu}^a
=
g_{\mu \lambda} \delta {e^\lambda}_a {u_\nu}^a.
\end{equation}
This can also be written as
\begin{equation}
{T^\lambda}_\nu
=
{e^\lambda}_a {u_\nu}^a
\rightarrow
\delta {T^\lambda}_\nu
=
\delta {e^\lambda}_a {u_\nu}^a.
\end{equation}
Inserting this back into the action for a graviational source (\ref{fundamental_variation_of_the_matter_action}) we have
\begin{subequations}
\begin{eqnarray}
I_M
& = &
\int d^4x \sqrt{- g} \, {T^\lambda}_\lambda
\\
& = &
\label{variation_of_matter_action_in_terms_of_vierbein_field}
\int d^4x \sqrt{- g} \, g^{\mu \nu} T_{\mu \nu}.
\end{eqnarray}
\end{subequations}

\subsection{Full gravitational action}

The variation of the full gravitational action is the sum of variations of the source-free action and gravitational action for matter
\begin{equation}
\label{total_variation_of_gravitational_action}
\delta I = \delta I_G + \delta I_M.
\end{equation}
Inserting (\ref{variation_of_source_free_gravitational_action}) and (\ref{variation_of_matter_action_in_terms_of_vierbein_field}) into  (\ref{total_variation_of_gravitational_action}) then gives
\begin{equation}
\delta I_G
 = 
 \!\!\!
\int d^4x \sqrt{- g}
\left[
\frac{1}{8 \pi G}
\left(
{R^\mu}_\lambda
-
\frac{1}{2} \delta^\mu_\lambda R
\right) {e_\mu}^a
+
{u_\lambda}^a
\right]
\delta {e^\lambda}_a.
\end{equation}
Therefore, with the requirement that $\delta I_G  =  0$, we obtain the equation of motion of the vierbein field
\begin{equation}
\left(
{R^\mu}_\lambda
-
\frac{1}{2} \delta^\mu_\lambda R
\right) {e_\mu}^a
=
- 8 \pi G {u_\lambda}^a.
\end{equation}
Multiplying through by $e_{\nu a}$
\begin{equation}
\left(
{R^\mu}_\lambda
-
\frac{1}{2} \delta^\mu_\lambda R
\right) {e_\mu}^a e_{\nu a}
=
- 8 \pi G e_{\nu a} {u_\lambda}^a
\end{equation}
gives the well known Einstein equation
\begin{equation}
R_{\nu \lambda}
-
\frac{1}{2} g_{\nu \lambda} R
=
- 8 \pi G \, T_{\nu \lambda}.
\end{equation}

\section{Einstein's action}
\label{Einsteins_action_in_vierbein_field_variables}

In this section, we review the derivation of the equation of motion of the metric field in the weak field approximation.  We start with a form of the Lagrangian density  presented in \cite{einstein-1928b} for the vierbein field theory.  Einstein's intention was the unification of electromagnetism with gravity.

With $h$ denoting the determinant of $|e_{\mu a}|$ ({\it i.e.} $h\equiv \sqrt{-g}$), the useful identity 
\begin{equation*}
\delta \sqrt{- g}
=
\frac{1}{2} \sqrt{- g} \,
g^{\mu \nu} \delta \, g_{\mu \nu}
\end{equation*}
can be rewritten strictly in terms of the vierbein field as follows
\begin{subequations}
\begin{eqnarray}
\delta h & = & \frac{1}{2} h \, g^{\mu \nu} \delta g_{\mu \nu} \\
& = & \frac{1}{2} h \, g^{\mu \nu} \delta ({e_\mu}^a {e_\nu}^b) \, \eta_{a b} \\
& = & \frac{1}{2} h \, g^{\mu \nu} \delta {e_\mu}^a {e_\nu}^b \eta_{a b}
+ \frac{1}{2} h \, g^{\mu \nu} {e_\mu}^a \delta {e_\nu}^b \eta_{a b} 
\qquad
\\
& = &  \frac{1}{2} h \, \delta {e_\mu}^a {e^\mu}_a
+ \frac{1}{2} h \, {e^\nu}_b \delta {e_\nu}^b \\
& = & h \, \delta {e_\mu}^a {e^\mu}_a.
\end{eqnarray}
\end{subequations}

\subsection{Lagrangian density form 1}

With the following definition
\begin{equation}
\Lambda^\nu_{\alpha \beta} \equiv \frac{1}{2} e^{\nu a} (\partial_\beta e_{\alpha a} - \partial_\alpha e_{\beta a}),
\end{equation}
 the first Lagrangian density that we consider is the following:
\begin{subequations}
\begin{eqnarray}
{\cal L} & = & h\, g^{\mu\nu} \; {\Lambda_\mu}^\alpha_\beta \; {\Lambda_\nu}^\beta_\alpha,
\\
\nonumber
& = & \frac{h}{4} g^{\mu\nu} e^{\alpha a} e^{\beta b}
(\partial_\beta e_{\mu a} - \partial_\mu e_{\beta a})
(\partial_\alpha e_{\nu b} - \partial_\nu e_{\alpha b}).
\qquad
\\
\end{eqnarray}
\end{subequations}
For a weak field, we have the following first-order expansion
\begin{equation}
e_{\mu a} = \delta_{\mu a} - k_{\mu a} \cdots .
\end{equation}
The lowest-order change (2nd order in $\delta h$) is
\begin{subequations}
\begin{eqnarray}
\nonumber
\delta {\cal L} 
& = &
\frac{h}{4} \eta^{\mu \nu} \delta^{\alpha a} \delta^{\beta b}
(\partial_\beta k_{\mu a} - \partial_\mu k_{\beta a})
(\partial_\alpha k_{\nu b} - \partial_\nu k_{\alpha b})
\qquad
\\
\\
& = &
\frac{h}{4} \eta^{\mu \nu}
(\partial_\beta {k_\mu}^\alpha - \partial_\mu {k_\beta}^\alpha)
(\partial_\alpha {k_\nu}^\beta - \partial_\nu {k_\alpha}^\beta)
\\
& = &
\frac{h}{4} \eta^{\mu \nu}
(\partial_\beta {k_\mu}^\alpha \partial_\alpha {k_\nu}^\beta 
- \partial_\beta {k_\mu}^\alpha \partial_\nu {k_\alpha}^\beta
\\
\nonumber
& &
- \, \partial_\mu {k_\beta}^\alpha \partial_\alpha {k_\nu}^\beta 
+ \partial_\mu {k_\beta}^\alpha \partial_\nu {k_\alpha}^\beta)
\\
& = &
\frac{h}{4}
\left(
\eta^{\mu \alpha}
\partial_\beta {k_\mu}^\nu \partial_\nu {k_\alpha}^\beta
- \eta^{\mu \nu} \partial_\beta {k_\mu}^\alpha \partial_\nu {k_\alpha}^\beta
\right.
\\
\nonumber
& &
\left.
- \, \eta^{\mu \alpha} \partial_\mu {k_\beta}^\nu \partial_\nu {k_\alpha}^\beta
+ \eta^{\mu \nu} \partial_\mu {k_\beta}^\alpha \partial_\nu {k_\alpha}^\beta
\right)
\\
& = &
\frac{h}{4}
\left(
- \eta^{\mu \alpha} \partial_\beta \partial_\nu {k_\mu}^\nu
+ \eta^{\mu \nu} \partial_\beta \partial_\nu {k_\mu}^\alpha
\right.
\\
& &
\left.
+ \, \eta^{\mu \alpha} \partial_\mu \partial_\nu {k_\beta}^\nu
- \eta^{\mu \nu} \partial_\mu \partial_\nu {k_\beta}^\alpha
\right)
{k_\alpha}^\beta.
\end{eqnarray}
\end{subequations}
So $\delta \int d^4x \,{\cal L} =0$ implies the equation of motion
\begin{subequations}
\begin{equation}
- \partial_\beta \partial_\nu k^{\alpha \nu} 
+ \partial_\beta \partial_\nu k^{\nu \alpha}
+ \partial^\alpha \partial_\nu {k_\beta}^\nu 
- \partial^2 {k_\beta}^\alpha 
= 0
\end{equation}
or
\begin{equation}
\label{equation_of_motion_Eq5}
\partial^2 {k_\beta}^\alpha 
- \partial_\mu \partial_\beta k^{\mu \alpha} 
+ \partial_\beta \partial_\mu k^{\alpha \mu}
- \partial_\mu \partial^\alpha {k_\beta}^\mu  
= 0.
\end{equation}
\end{subequations}
The above equation of motion (\ref{equation_of_motion_Eq5}) is identical to Eq.~(5) in Einstein's second paper.

\subsection{Lagrangian density form 2}

Now the second Lagrangian density we consider is the following:
\begin{subequations}
\begin{eqnarray}
{\cal L} 
\!\!\!
& = &
h \, g_{\mu \nu} g^{\alpha \sigma} g^{\beta \tau} 
\Lambda^\mu_{\alpha \beta} \Lambda^\nu_{\sigma \tau}
\\
\nonumber
& = &
\frac{h}{4} g_{\mu \nu} g^{\alpha \sigma} g^{\beta \tau}
e^{\mu a} e^{\nu b}
\left(
\partial_\beta e_{\alpha a}
- \partial_\alpha e_{\beta a}
\right)
\left(
\partial_\tau e_{\sigma b}
- \partial_\sigma e_{\tau b}
\right).
\\
\end{eqnarray}
\end{subequations}
We will see this leads to the same equation of motion that we got from the first form of the  Lagrangian density.
The lowest-order change is
\begin{subequations}
\begin{eqnarray}
\nonumber
\delta{\cal L}
\!\!\!
& = &
\!\!\!
\frac{h}{4} \eta_{\mu \nu} \eta^{\alpha \sigma} \eta^{\beta \tau}
\delta^{\mu a} \delta^{\nu b}
\left(
\partial_\beta k_{\alpha a}
- \partial_\alpha k_{\beta a}
\right)
\left(
\partial_\tau e_{\sigma b}
- \partial_\sigma e_{\tau b}
\right)
\\
& &
\\
& = &
\frac{h}{4} \eta^{\alpha \sigma} \eta^{\beta \tau}
\left(
\partial_\beta k_{\alpha \nu}
- \partial_\alpha k_{\beta \nu}
\right)
\left(
\partial_\tau {k_\sigma}^\nu
- \partial_\sigma {k_\tau}^\nu
\right)
\\
\label{Amazing_quadratic_Lagrangian_density}
& = &
\frac{h}{4}
\left(
\partial_\beta k_{\alpha \nu}
- \partial_\alpha k_{\beta \nu}
\right)
\left(
\partial^\beta k^{\alpha \nu}
- \partial^\alpha k^{\beta \nu}
\right)
\\
\nonumber
& = &
\frac{h}{4}
\left(
\partial_\beta k_{\alpha \nu}
\partial^\beta k^{\alpha \nu}
- \partial_\beta k_{\alpha \nu}
\partial^\alpha k^{\beta \nu}
- \partial_\alpha k_{\beta \nu}
\partial^\beta k^{\alpha \nu}
\right.
\\
& &
\left.
+ \, \partial_\alpha k_{\beta \nu}
\partial^\alpha k^{\beta \nu}
\right)
\\
\nonumber
& = &
\frac{h}{4}
\left(
\partial_\beta k_{\alpha \nu}
\partial^\beta k^{\alpha \nu}
- \partial_\beta k_{\alpha \nu}
\partial^\alpha k^{\beta \nu}
- \partial_\beta k_{\alpha \nu}
\partial^\alpha k^{\beta \nu}
\right.
\\
& &
\left.
+ \, \partial_\beta k_{\alpha \nu}
\partial^\beta k^{\alpha \nu}
\right)
\\
& = &
\frac{h}{2}
\left(
- \partial^2 k^{\alpha \nu}
+ \partial_\beta \partial^\alpha k^{\beta \nu}
\right)
k_{\alpha \nu}.
\end{eqnarray}
\end{subequations}
This implies the following:
\begin{subequations}
\begin{equation}
\label{equation_of_motion_form_2_indices_up}
\partial^2 k^{\alpha \nu}
- \partial_\beta \partial^\alpha k^{\beta \nu}
=
0
\end{equation}
or
\begin{equation}
\label{equation_of_motion_form_2}
\partial^2 {k_\beta}^\alpha
- \partial_\mu \partial_\beta
k^{\mu \alpha}
=
0.
\end{equation}
\end{subequations}
These are the first two terms in Einstein's Eq.~(5).  

Notice that we started with a Lagrangian density with the usual quadratic form in the field strength of the form (\ref{Amazing_quadratic_Lagrangian_density}), which is 
\begin{equation}
{\cal L} = \frac{h}{4} F_{\alpha\beta\nu}F^{\alpha\beta\nu},
\end{equation}
where the field strength is
\begin{equation}
F^{\alpha\beta\nu}=
 \partial^\alpha k^{\beta \nu}
-\partial^\beta k^{\alpha \nu}.
\end{equation}
If we had varied the action with respect to $k^{\beta \nu}$, then we would have obtained the same equation of motion (\ref{equation_of_motion_form_2}).

\subsection{First-order fluctuation in the metric tensor}

The metric tensor expressed in terms of the vierbein field is
\begin{equation}
g_{\alpha \beta}
=
{e_\alpha}^a e_{\beta a}
=
\left(
\delta^a_\alpha + {k_\alpha}^a
\right)
\left(
\delta_{\beta a}
+ k_{\beta a}
\right).
\end{equation}
So the first order fluctuation of the metric tensor field is the symmetric tensor
\begin{equation}
\label{symmetry_first_order_fluctuation_of_metric_tensor}
\overline{g_{\alpha \beta}}
\equiv
g_{\alpha \beta} - \delta_{\alpha \beta}
=
k_{\alpha \beta}
+ k_{\beta \alpha} \cdots .
\end{equation}
We define the electromagnetic four-vector by contracting the field strength tensor
\begin{equation}
\varphi_\mu
\equiv
\Lambda_{\mu \alpha}^\alpha
= \frac{1}{2} e^{\alpha a}
\left(
\partial_\alpha e_{\mu a}
- \partial_\mu e_{\alpha a}
\right).
\end{equation}
This implies
\begin{equation}
\varphi_\mu
=
\frac{1}{2} \delta^{\alpha a}
\left(
\partial_\alpha k_{\mu a}
- \partial_\mu k_{\alpha a}
\right),
\end{equation}
so we arrive at
\begin{equation}
\label{Einstein's_equation_2a}
2 \varphi_\mu
=
\partial_\alpha {k_\mu}^\alpha
- \partial_\mu {k_\alpha}^\alpha.
\end{equation}

\subsection{Field equation in the weak field limit}

The equation of motion for the fluctuation of the metric tensor from the first form of the  Lagrangian density is obtained by adding 
(\ref{equation_of_motion_Eq5}) to itself but with $\alpha$ and $\beta$ exchanged:
\begin{equation}
\begin{split}
&
\partial^2 k_{\beta \alpha}
- \partial^\mu \partial_\beta k_{\mu \alpha}
+ \partial_\beta \partial^\mu k_{\alpha \mu}
- \partial^\mu \partial_\alpha k_{\beta \mu}
\\
+ &
\partial^2 k_{\alpha \beta}
- \partial^\mu \partial_\alpha k_{\mu \beta}
+ \partial_\alpha \partial^\mu k_{\beta \mu}
- \partial^\mu \partial_\beta k_{\alpha \mu}
= 0, \quad
\end{split}
\end{equation}
which has a cancellation of four terms leaving
\begin{equation}
\label{first_summation}
\partial^2 \overline{g_{\alpha \beta}}
- \partial^\mu \partial_\alpha k_{\mu \beta}
- \partial^\mu \partial_\beta k_{\mu \alpha}
= 0.
\end{equation}
Similarly, we arrive at the same result starting with the equation of motion for the fluctuation of the metric tensor obtained from the second form of the  Lagrangian density, again by adding 
(\ref{equation_of_motion_form_2_indices_up}) to itself but with $\alpha$ and $\beta$ exchanged:
\begin{equation}
\begin{split}
&
\partial^2 k_{\beta \alpha}
- \partial^\mu \partial_\beta k_{\mu \alpha}
\\
+&
\, \partial^2 k_{\alpha \beta}
- \partial^\mu \partial_\alpha k_{\mu \beta}
=
0.
\end{split}
\end{equation}
In this case, the sum 
 is exactly the same as what we just obtained in (\ref{first_summation})  but with no cancellation of terms
\begin{equation}
\label{base_equation_of_motion}
\partial^2 \overline{g_{\alpha \beta}}
- \partial^\mu \partial_\alpha k_{\mu \beta}
- \partial^\mu \partial_\beta k_{\mu \alpha}
= 0.
\end{equation}
Using 
(\ref{Einstein's_equation_2a}) above, just with relabeled indices,
\begin{equation}
\label{Einstein's_equation_2a_switched_indices}
2 \varphi_\alpha
=
\partial_\mu {k_\alpha}^\mu
- \partial_\alpha {k_\mu}^\mu.
\end{equation}
Taking derivatives of (\ref{Einstein's_equation_2a_switched_indices}) we have ancillary equations of motion:
\begin{subequations}
\label{ancillary_equations_of_motion}
\begin{equation}
- \partial_\mu \partial_\beta
{k_\alpha}^\mu
+ \partial_\alpha \partial_\beta
{k_\mu}^\mu
=
- 2 \partial_\beta \varphi_\alpha
\end{equation}
and
\begin{equation}
- \partial_\mu \partial_\alpha
{k_\beta}^\mu
+ \partial_\alpha \partial_\beta
{k_\mu}^\mu
=
- 2 \partial_\alpha \varphi_\beta.
\end{equation}
\end{subequations}
Adding the ancilla (\ref{ancillary_equations_of_motion}) to our equation of motion (\ref{base_equation_of_motion}) gives
\begin{eqnarray}
\nonumber
-\partial^{2}\overline{g_{\alpha\beta}} + \partial^\mu\partial_\alpha(k_{\mu\beta}&+&k_{\beta\mu}) 
+  \partial^\mu\partial_\beta(k_{\mu\alpha}+k_{\alpha\mu}) 
\\
&  -&2 \partial_{\alpha}  \partial_{\beta}  k_\mu^\mu
\nonumber =
2(\partial_{\beta}\varphi_\alpha+ \partial_{\alpha}\varphi_\beta ).
\\
\end{eqnarray}
Then making use of (\ref{symmetry_first_order_fluctuation_of_metric_tensor}) this can be written in terms of the symmetric first-order fluctuation of the metric tensor field
\begin{eqnarray}
\label{equation_of_motion_Einstein_action_form_2}
\nonumber
\left.
\frac{1}{2}
\middle(
- \partial^2 \overline{g_{\alpha \beta}}
+ \partial^\mu \partial_\alpha
\overline{g_{\mu \beta}}
\right.
&+&
\left.
 \partial^\mu \partial_\beta \overline{g_{\mu \alpha}}
- \partial_\alpha \partial_\beta \overline{{{g}_\mu}^\mu}
\frac{ }{ }
\right) 
\\
& = &
\partial_\beta \varphi_\alpha
+ \partial_\alpha \varphi_\beta.
\end{eqnarray}
This result is the same as Eq.~(7) in Einstein's second paper.
In the case of the vanishing of  $\phi_\alpha$, (\ref{equation_of_motion_Einstein_action_form_2}) agrees to first order with the equation of General Relativity
\begin{equation}
R_{\alpha\beta}=0.
\end{equation}
 Thus, Einstein's action expressed explicitly in terms of the vierbein field reproduces the law of the pure gravitational field in weak field limit.

\section{Relativistic chiral matter in curved space}
\label{relativistic_chiral_matter_in_curved_space}

\subsection{Invariance in flat space}

The external Lorentz transformations,
$\Lambda$ that act on 4-vectors, commute with the internal Lorentz
transformations, $U(\Lambda)$ that act on  spinor wave functions, {\it i.e.}
\begin{equation}
\label{eq:Lambda_D_commute}
[{\Lambda^{\mu}}_{\nu},U(\Lambda)]=0.
\end{equation}
Note that we keep the indices on $U(\Lambda)$ suppressed, just as
we keep the indices of the Dirac matrices and the component indices
of $\psi$ suppressed as is conventional when writing matrix multiplication.
Only the exterior spacetime indices are explicitly written out. With
this convention, the Lorentz transformation of a Dirac gamma matrix
is expressed as follows:
\begin{equation}
\label{eq:Lambda_D_similarity_transform}
U(\Lambda)^{-1}\gamma^{\mu}U(\Lambda)={\Lambda^{\mu}}_{\sigma}\gamma^{\sigma}.
\end{equation}
The invariance of the Dirac equation in flat space under a Lorentz transformation is well known \cite{Peskin_Schroeder_2004}:
\begin{widetext}
\begin{subequations}
\begin{eqnarray}
\left[i\gamma^{\mu}\partial_{\mu}-m\right]
\psi(x)
& \stackrel{\text{\tiny LLT}}{\longrightarrow}  &
\left[
i\gamma^{\mu}
{\left(
\Lambda^{-1}
\right)^{\nu}}_{\mu}
\partial_{\nu}-m\right]U(\Lambda)\psi\left(\Lambda^{-1}x\right)
\\
 & = & U(\Lambda)U(\Lambda)^{-1}\left[i\gamma^{\mu}
 {\left(\Lambda^{-1}\right)^{\nu}}_{\mu}\partial_{\nu}-m\right]
 U(\Lambda)\psi\left(\Lambda^{-1}x\right)
 \\
 & \stackrel{(\ref{eq:Lambda_D_commute})}{=} & 
 U(\Lambda)\left[i\, U(\Lambda)^{-1}\gamma^{\mu}U(\Lambda){\left(\Lambda^{-1}\right)^{\nu}}_{\mu}\partial_{\nu}-m\right]\psi\left(\Lambda^{-1}x\right)
 \\
 & \stackrel{(\ref{eq:Lambda_D_similarity_transform})}{=}  & U(\Lambda)\left[i\,\Lambda_{\sigma}^{\mu}\gamma^{\sigma}{\left(\Lambda^{-1}\right)^{\nu}}_{\mu}\partial_{\nu}-m\right]\psi\left(\Lambda^{-1}x\right)\\
 & = & U(\Lambda)\left[i\,\Lambda_{\sigma}^{\mu}
 {\left(\Lambda^{-1}\right)^{\nu}}_{\mu}
 \gamma^{\sigma}\partial_{\nu}-m\right]\psi\left(\Lambda^{-1}x\right)\\
 & = & U(\Lambda)
 \left[i\,\delta_{\sigma}^{\nu}\,\gamma^{\sigma}\partial_{\nu}-m\right]\psi
 \left(\Lambda^{-1}x\right)\\
 & = & U(\Lambda)\left[i\,\gamma^{\nu}\partial_{\nu}-m\right]\psi\left(\Lambda^{-1}x\right).
 \end{eqnarray}
\end{subequations}
\end{widetext}

\subsection{Invariance  in curved space}
\label{Invariance_in_curved_space}

Switching to a compact notation for the interior Lorentz transformation, $\Lambda_{\frac{1}{2}}\equiv U(\Lambda)$, (\ref{eq:Lambda_D_similarity_transform})  is
\begin{equation}
\label{eq:Lorentz_interior_similarity_transformation}
\Lambda_{- \frac{1}{2}}
\gamma^\mu
\Lambda_{\frac{1}{2}}
=
{\Lambda^\mu}_\sigma
\gamma^\sigma,
\end{equation}
where I put a minus on the subscript to indicate the inverse transformation, {\it i.e.} $\Lambda_{-\frac{1}{2}}\equiv U(\Lambda)^{-1}$.
Of course, exterior Lorentz transformations can be used as a similarity transformation on the Dirac matrices
\begin{equation}
\label{eq:Lorentz_exterior_similarity_transformation}
{\Lambda^\mu}_\sigma
\gamma^\sigma
{\left(
\Lambda^{-1}
\right)^\nu}_\mu
=
\gamma^\nu.
\end{equation}
Below we will need the following identity:
%
\begin{subequations}
\begin{eqnarray}
{\Lambda^\mu}_\lambda
{e^\lambda}_a \gamma^a
{{(\Lambda^{-1})}^\nu}_\mu
\Lambda_{\frac{1}{2}}
\nonumber
& \stackrel{(\ref{eq:Lambda_D_commute})}{=} &
\Lambda_{\frac{1}{2}}
\Lambda_{-\frac{1}{2}}
\left(
{\Lambda^\mu}_\lambda
{e^\lambda}_a \gamma^a
\right)
\Lambda_{\frac{1}{2}}
{{(\Lambda^{-1})}^\nu}_\mu
\\
& &
\\
\nonumber
& \stackrel{(\ref{eq:Lorentz_interior_similarity_transformation})}{=}  &
\Lambda_{\frac{1}{2}}
{\Lambda^\mu}_\sigma
\left(
{\Lambda^\sigma}_\lambda
{e^\lambda}_a \gamma^a
\right)
{{(\Lambda^{-1})}^\nu}_\mu
\\
\\
\label{identity_Lorentz_exterior_veirbein_gamma_inverse_spinor}
& \stackrel{(\ref{eq:Lorentz_exterior_similarity_transformation})}{=}  &
\Lambda_{\frac{1}{2}}
{\Lambda^\nu}_\lambda
{e^\lambda}_a \gamma^a.
\end{eqnarray}
\end{subequations}
%
We require the Dirac equation in curved space be invariant under Lorentz transformation when the curvature of space causes a correction $\Gamma_\mu$.  That is, we require
\begin{subequations}
\begin{eqnarray}
\nonumber
& 
{e^\mu}_a
&
 \gamma^a
\left(
\partial_\mu + \Gamma_\mu
\right) \psi(x)
\\
& 
\stackrel{\text{\tiny LLT}}{\longrightarrow} 
&
{\Lambda^\mu}_\lambda
{e^\lambda}_a \gamma^a
{{(\Lambda^{-1})}^\nu}_\mu
\left(
\partial_\nu + \Gamma'_\nu
\right)
\Lambda_{\frac{1}{2}}
\psi(\Lambda^{-1} x)
\\
\nonumber
&  =&
{\Lambda^\mu}_\lambda
{e^\lambda}_a \gamma^a
{{(\Lambda^{-1})}^\nu}_\mu
\Lambda_{\frac{1}{2}}
\left(
\partial_\nu
+ 
\Lambda_{-\frac{1}{2}}
\Gamma'_\nu \Lambda_{\frac{1}{2}}
\right)
\psi(\Lambda^{-1} x)
\\
& &
+ \, {\Lambda^\mu}_\lambda
{e^\lambda}_a \gamma^a
{\left(
\Lambda^{-1}
\right)^\nu}_\mu
\left(
\partial_\nu
\,
\Lambda_{\frac{1}{2}}
\right)
\psi(\Lambda^{-1} x)
\\
\nonumber
& \stackrel{(\ref{identity_Lorentz_exterior_veirbein_gamma_inverse_spinor})}{=} &
\Lambda_{\frac{1}{2}}
{\Lambda^\nu}_\lambda
{e^\lambda}_a \gamma^a
\left[
\left(
\partial_\nu
+
\Lambda_{-\frac{1}{2}}
\Gamma'_\nu
\Lambda_{\frac{1}{2}}
\right)
\psi(\Lambda^{-1} x)
\right.
\\
& &
\left.
+
\, \Lambda_{-\frac{1}{2}}
\left(
\partial_\nu
\,
\Lambda_{\frac{1}{2}}
\right)
\psi(\Lambda^{-1} x)
\right]
\\
\nonumber
& = &
\Lambda_{\frac{1}{2}}
{\Lambda^\nu}_\lambda
{e^\lambda}_a \gamma^a
\left[
\left(
\partial_\nu + \Gamma_\nu
\right)
\right.
\\
\nonumber
& &
-
\, 
\underbrace{\Gamma_\nu
+
\Lambda_{-\frac{1}{2}}
\Gamma'_\nu
\Lambda_{\frac{1}{2}}
+
\Lambda_{-\frac{1}{2}}
\partial_\nu
\left(
\Lambda_{\frac{1}{2}}
\right)}_{=0}
\left]
\psi(\Lambda^{-1} x).
\right.
\qquad
\\
\end{eqnarray}
\end{subequations}
In the last line we added and subtracted $\Gamma_\nu$.  To achieve invariance,  the last three terms in the square brackets must vanish.
Thus we find the form of the local ``gauge'' transformation requires the correction field to transform as follows:
\begin{subequations}
\begin{equation}
- \Gamma_\nu
+
\Lambda_{-\frac{1}{2}}
\Gamma'_\nu
\Lambda_{\frac{1}{2}}
+
\Lambda_{-\frac{1}{2}}
\partial_\nu
\left(
\Lambda_{\frac{1}{2}}
\right)
= 0
\end{equation}
or
\begin{equation}
\label{Gamma_pseudo_gauge_transformation}
\Gamma'_\nu
=
\Lambda_{\frac{1}{2}}
\Gamma_\nu
\Lambda_{-\frac{1}{2}}
-
\partial_\nu
\left(
\Lambda_{\frac{1}{2}}
\right)
\Lambda_{-\frac{1}{2}}.
\end{equation}
\end{subequations}
 Therefore, the Dirac equation in curved space
\begin{equation}
i\gamma^a {e^\mu}_a(x) {\cal D}_\mu \, \psi - m\, \psi=0
\end{equation}
is invariant under a Lorentz transformation provided the generalized derivative that we use is
\begin{equation}
 {\cal D}_\mu=\partial_\mu+\Gamma_\mu,
\end{equation}
where $\Gamma_\mu$ transforms according to (\ref{Gamma_pseudo_gauge_transformation}).  This is analogous to a gauge correction; however, in this case $\Gamma_\mu$ is not a vector potential field.

\subsection{Covariant derivative of a spinor field}

\begin{figure}[htbp]
\begin{center}
\setlength{\unitlength}{.5mm}
\begin{picture}(65,45)

{\color{blue_ud}\qbezier(-5,-2.5)(20,10)(55,-1.5)}

\put(0,0){\vector(-1,3){9}}
{\color{red_ud}\put(50,0){\vector(-1,3){9}}}
{\color{blue_ud}\put(48,0){\vector(1,3){9}}}
{\color{red_ud}\put(-4,0){\line(1,0){50}}}
\put(54.5,27){\vector(-1,0){17.5}}

\put(-6,-8){$x^\alpha$}
\put(39,-8){$x^\alpha + \delta x^\alpha$}

\put(-32,14){$X^{\alpha_\parallel} (x)$}
{\color{blue_ud}\put(52,14){$X^{\alpha_\parallel}(x + \delta x)$}}
\put(20,30){$\Gamma^{\alpha_\parallel}_{\beta \gamma}(x)
X^{\beta}(x) \delta x^\gamma$}

\put(46,0){\circle*{2}}
\put(-5,0){\circle*{2}}

\end{picture}
\vspace{0.2in}
\end{center}
\caption{\label{parallel_transport_in_curved_space}\footnotesize Depiction of the case of an otherwise constant field distorted by curved space.  The field value $X^\alpha(x)$ is parallel transported along the curved manifold ({\color{blue}blue curve}) by the distance $\delta x^\alpha$ going from point $x^\alpha$ to $x^\alpha+\delta x^\alpha$.}
\end{figure}
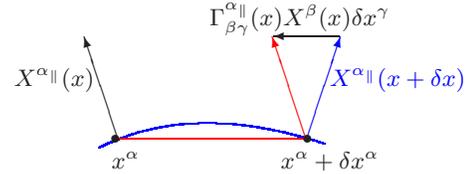

The Lorentz transformation for a spinor field is
\begin{equation}
\label{Lorentz_transformation_spinor_field}
\Lambda_{\frac{1}{2}}
=
1 + \frac{1}{2} \lambda_{a b} \, S^{a b},
\end{equation}
where the generator of the transformation is anti-symmetric $S^{a b} = - S^{b a}$.  The generator satisfies the following commutator
\begin{equation}
[S^{h k}, S^{i j}]
=
\eta^{h j} S^{k i}
+
\eta^{k i} S^{h j}
-
\eta^{h i} S^{k j}
-
\eta^{k j} S^{h i}.
\end{equation}
Thus, the local Lorentz transformations (LLT) of a Lorentz 4-vector,  $x^a$ say, and a Dirac 4-spinor, $\psi$ say, are respectively:
\begin{equation}
\text{LLT:}\qquad x^a \rightarrow x'^a
= {\Lambda^a}_b \, x^b
\end{equation}
and
\begin{equation}
\text{LLT:}\qquad \psi \rightarrow \psi'
=
\Lambda_{\frac{1}{2}} \psi.\quad
\end{equation}
The covariant derivative of a 4-vector is
\begin{equation}
\nabla_\gamma X^\alpha
=
\partial_\gamma X^\alpha
+
\Gamma^\alpha_{\beta \gamma} X^\beta,
\end{equation}
and the  4-vector at the nearby location is changed by the curvature of the manifold. So we write it in terms of the original 4-vector with a correction
\begin{equation}
X^{\alpha_\parallel}
(x + \delta x^\alpha)
=
X^{\alpha_\parallel}(x)
-
\Gamma^{\alpha_\parallel}_{\beta \gamma}(x)
X^{\beta}(x) \delta x^\gamma,
\end{equation}
as depicted in Fig.~\ref{parallel_transport_in_curved_space}.

Likewise, the correction to  the vierbein field due the curvature of space is
\begin{equation}
\label{parallel_transport_correction_to_vierbein_field}
{e^{\mu_\parallel}}_k
(x + \delta x^\alpha)
=
{e^{\mu_\parallel}}_k(x)
-
\Gamma^{\mu_\parallel}_{\beta \alpha}(x)
{e^\beta}_k(x) \delta x^\alpha.
\end{equation}
The Lorentz transformation of a 2-rank tensor field is
\begin{equation}
{\Lambda^a}_{a'}
{\Lambda^b}_{b'}
\eta_{a b}
=
\eta_{a' b'}.
\end{equation}
Moreover, the Lorentz transformation is invertible
\begin{equation}
{\Lambda^i}_a 
{\Lambda_j}^a
=
\delta^i_j
=
{\Lambda^a}_j
{\Lambda_a}^i,
\end{equation}
where the inverse is obtained by exchanging index labels, changing covariant indices to contravariant indices and contravariant to covariant.
In the case of infinitesimal transformations we have
\begin{equation}
\label{infinitesimal_Lorentz_transformation}
{\Lambda^i}_j(x)
=
\delta^i_j + {\lambda^i}_j(x),
\end{equation}
where
\begin{equation}
0 = \lambda_{i j} + \lambda_{j i}
= \lambda^{i j} + \lambda^{j i}.
\end{equation}
%
%
%
%

Lorentz and inverse Lorentz transformations of the vierbein fields are
\begin{equation}
\bar{e}^\mu_{\, \, h'}(x)
=
{\Lambda_{h'}}^a(x)
{e^\mu}_a(x)
\end{equation}
and
\begin{equation}
\label{inverse_Lorentz_transformation_of_vierbein}
{e^\mu}_{h}(x)
=
{\Lambda^{a'}}_h(x)
{{\bar{e}}^\mu}_{\;\;a'}(x),  
\end{equation}
where temporarily I am putting a bar over the transformed vierbein field as a visual aid.  Since the vierbein field is invertible, we can express the Lorentz tranformation directly in terms of the vierbeins themselves
\begin{equation}
\label{Lorentz_transformation_as_product_of_vierbeins}
{\bar{e}_\mu}^{\, \, \, a'}(x)
{e^\mu}_h(x)
=
{\Lambda^{a'}}_h(x).
\end{equation}
Now, we transport the Lorentz transformation tensor itself.  The L.H.S. of (\ref{Lorentz_transformation_as_product_of_vierbeins}) has two upper indices, the Latin index $a'$ and the Greek index $\mu$, and we choose to use the upper indices to connect the Lorentz transformation tensor between neighboring points.    These indices are treated differently: a Taylor expansion can be used to connect a quantity in its Latin non-coordinate index at one point to a neighboring point, but the affine connection must be used for the Greek coordinate index.  Thus, we have
\begin{subequations}
\begin{eqnarray}
{\Lambda^{h'}}_k(x + \delta x^\alpha)
& = &
{\bar{e}_{\mu_\parallel}}^{\quad h'} (x + \delta x^\alpha)
{e^{\mu_\parallel}}_k (x + \delta x^\alpha) 
\\
\nonumber
&=&
\left(
{\bar{e}_\mu}^{\, \, \, h}(x)
+
\frac{\partial{\bar{e}_\mu}^{\, \, \, h}}{\partial x^\alpha}
\delta x^\alpha
\right)
{e^{\mu_\parallel}}_k (x + \delta x^\alpha) 
\\
\\
& \stackrel{(\ref{parallel_transport_correction_to_vierbein_field})}{=} &
\left(
{\bar{e}_\mu}^{\, \, \, h}(x)
+
\frac{\partial{\bar{e}_\mu}^{\, \, \, h}}{\partial x^\alpha}
\delta x^\alpha
\right)
\\
\nonumber
& &
\times\left(
{e^\mu}_k
-
\Gamma^\mu_{\beta \alpha}(x)
{e^\beta}_k(x) \delta x^\alpha
\right)
\\
\nonumber
& = &
\delta^h_k
+
\frac{\partial {\bar{e}_\mu}^{\, \, \, h}}{\partial x^\alpha}
\delta x^\alpha {e^\mu}_k
-
\Gamma^\mu_{\beta \alpha} {e^\beta}_k
\delta x^\alpha {\bar{e}_\mu}^{\, \, \, h}
\\
& &
\\
\nonumber
& = &
\delta^h_k
+
\left(
\frac{\partial {\bar{e}_\mu}^{\, \, \, h}}{\partial x^\alpha}
\delta^\mu_\beta
-
\Gamma^\mu_{\beta \alpha}
{\bar{e}_\mu}^{\, \, \, h}
\right)
{e^\beta}_k \delta x^\alpha
\\
\\
\nonumber
& = &
\delta^h_k
+
\left(
{e^\mu}_k \partial_\alpha
{\bar{e}_\mu}^{\, \, \, h}
-
\Gamma^\mu_{\beta \alpha}
{\bar{e}_\mu}^{\, \, \, h}
{e^\beta}_k
\right)
\delta x^\alpha
\\
\\
& = &
\delta^h_k
-
{{\omega_\alpha}^h}_k
\delta x^\alpha,
\end{eqnarray}
\end{subequations}
where the spin connection
\begin{equation}
{{\omega_\alpha}^h}_k=-{e^\mu}_k \partial_\alpha
{\bar{e}_\mu}^{\, \, \, h}
+
\Gamma^\mu_{\beta \alpha}
{\bar{e}_\mu}^{\, \, \, h}
{e^\beta}_k
\end{equation}
is seen to have the physical interpretation of generalizing the infinitesimal transformation (\ref{infinitesimal_Lorentz_transformation}) to the case of infinitesimal transport in curved space.
Relabeling indices, we have
\begin{subequations}
\begin{eqnarray}
{{\omega_\mu}^a}_b
& = &
- {e^\nu}_b \partial_\mu {e_\nu}^a
+
\Gamma^\sigma_{\mu \nu}
{e_\sigma}^a {e^\nu}_b
\\
\label{spin_connection_tetrad_postulate}
& = &
- {e^\nu}_b
\left(
\partial_\mu {e_\nu}^a
-
\Gamma^\sigma_{\mu \nu} {e_\sigma}^a
\right)
\\
& = &
- {e^\nu}_b \nabla_\mu {e_\nu}^a,
\end{eqnarray}
\end{subequations}
where here the covariant derivative of the vierbien 4-vector is not zero.\footnote{
Remember the Tetrad postulate that we previously derived
\begin{equation*}
\nabla_\mu {e_\nu}^a
=
\partial_\mu {e_\nu}^a
-
{e_\sigma}^a
\Gamma^\sigma_{\mu \nu}
+
{{\omega_\mu}^a}_b
{e_\nu}^b
= 0
\end{equation*}
is exactly (\ref{spin_connection_tetrad_postulate}).
}
Writing the Lorentz transformation in the usual infinitesimal form
\begin{equation}
{\Lambda^h}_k
=
\delta^h_k
+
{\lambda^h}_k
\end{equation}
implies
\begin{subequations}
\begin{eqnarray}
{\lambda^h}_k
& = &
- {{\omega_\alpha}^h}_k
\, \delta x^\alpha
\\
& = &
{e^\nu}_k
\left(
\nabla_\alpha {e_\nu}^h
\right)
\delta x^\alpha
\end{eqnarray}
or
\begin{equation}
\lambda_{h k}
=
{e^\beta}_k
\left(
\nabla_\alpha e_{\beta h}
\right)
\delta x^\alpha.
\end{equation}
\end{subequations}
Using (\ref{Lorentz_transformation_spinor_field}), the Lorentz transformation of the spinor field is
\begin{equation}
\Lambda_{\frac{1}{2}} \psi
=
\left(
1 + \frac{1}{2} \lambda_{h k} \, S^{h k}
\right)
\psi
=
\psi + \delta \psi.
\end{equation}
This implies the change of the spinor is
\begin{subequations}
\begin{eqnarray}
\delta \psi
& = &
\frac{1}{2} {e^\beta}_k
\left(
\nabla_\alpha e_{\beta h}
\right)
\delta x^\alpha S^{h k} \psi
\\
& = &
\Gamma_\alpha \psi \, \delta x^\alpha,
\end{eqnarray}
\end{subequations}
where the correction to the spinor field is found to be
\begin{subequations}
\begin{eqnarray}
\Gamma_\alpha
&=&
\frac{1}{2} {e^\beta}_k
\left(
\nabla_\alpha e_{\beta h}
\right)
S^{h k}
\\
&=&
\frac{1}{2} {e^\beta}_k
\left(
\partial_\alpha e_{\beta h}
-
\Gamma_{\mu\beta}^\sigma e_{\sigma h}
\right)
S^{h k}
\\
&=&
\label{Gamma_mu_line_3}
\frac{1}{2} {e^\beta}_k
\left(
\partial_\alpha e_{\beta h}
\right)
S^{h k}
-
\frac{1}{2} 
\Gamma_{\mu\beta}^\sigma (e_{\sigma h}{e^\beta}_k )
S^{h k}
\qquad
\\
&=&
\frac{1}{2} {e^\beta}_k
\left(
\partial_\alpha e_{\beta h}
\right)
S^{h k},
\end{eqnarray}
\end{subequations}
where the last term in (\ref{Gamma_mu_line_3}) vanishes because $e_{\sigma h}{e^\beta}_k$ is symmetric whereas $S^{h k}$ is anti-symmetric  in the indices $h$ and $k$.
Thus, we have derived the form of the covariant derivative of the spinor wave function
\begin{subequations}
\begin{eqnarray}
{\cal D}_\mu\psi
& = &
\partial_\mu \psi
+
\Gamma_\mu \psi
\\
& = &
\left(
\partial_\mu
+
\frac{1}{2} {e^\beta}_k
\nabla_\mu e_{\beta h}
\,
S^{h k}
\right)
\psi
\\
& = &
\left(
\partial_\mu
+
\frac{1}{2} {e^\beta}_k
\,
\partial_\mu e_{\beta h}
\,
S^{h k}
\right)
\psi.
\end{eqnarray}
\end{subequations}
This is the generalized derivative that is needed to correctly differentiate a Dirac 4-spinor field in curved space.

\section{Conclusion}

A detailed derivation of the Einstein equation from the least action principle and a derivation of the relativistic Dirac equation in curved space from considerations of invariance with respect to Lorentz transformations have been presented.   
The field theory approach that was presented herein relied on a factored decomposition of the metric tensor field in terms of a product of vierbein fields that Einstein introduced in 1928.  In this sense, the vierbein field is considered the square root of the metric tensor.  The motivation for this decomposition follows naturally from the anti-commutator
\begin{equation}
\label{fundamental_anticommutator_identity}
\{ {e^\mu}_a(x) \gamma^a, {e^\nu}_b(x) \gamma^b \} = 2g^{\mu\nu}(x), 
\end{equation}
where $\gamma^a$ are the Dirac matrices.  Dirac originally discovered an aspect of this important identity when he successfully attempted to write down a linear quantum wave equation that when squared gives the well known Klein-Gordon equation.  Thus, dealing with relativistic quantum mechanics in flat space, Dirac wrote this identity as
\begin{subequations}
\begin{equation}
\label{fundamental_anticommutator_identity_part_1}
\{ \gamma^a,  \gamma^b \} = 2\eta^{ab}, 
\end{equation}
where $\eta = \text{diag}(1,-1,-1,-1)$. Einstein had the brilliant insight to write the part of the  identity that depends on the spacetime curvature as
\begin{equation}
\label{fundamental_anticommutator_identity_part_2}
 {e^\mu}_a(x)  {e^\nu}_b(x) \eta^{ab} = g^{\mu\nu}(x).
\end{equation}
\end{subequations}
Combining (\ref{fundamental_anticommutator_identity_part_1}) and (\ref{fundamental_anticommutator_identity_part_2}) into (\ref{fundamental_anticommutator_identity}) is essential to correctly develop a relativistic quantum field theory in curved space.  However, (\ref{fundamental_anticommutator_identity_part_2}) in its own right is a sufficient point of departure if one seeks to simply derive the Einstein equation capturing the dynamical behavior of spacetime.

\section{Acknowledgements}

I would like to thank Carl Carlson for checking the derivations presented above.  
I would like to thank Hans C. von Baeyer for his help searching for past English translations of the 1928 Einstein manuscripts (of which none were found) and his consequent willingness to translate the German text into English.

\newpage

\lhead{} 
\rhead{} 
\thispagestyle{plain} 

\begin{appendices}

%

\onecolumngrid

The following two manuscripts, translated here in English by H.C. von Baeyer and into \LaTeX ~by the author, originally appeared in German in Sitzungsberichte der Preussischen Akademie der Wissenschaften, Physikalisch-Mathematische Klasse in the summer of 1928.

\section*{Einstein's 1928 manuscript on distant parallelism}

\newpage
\begin{center}
\mbox{
{\bf\large Riemann geometry with preservation of the concept of distant parallelism}
}
\\
\vspace{1em}
\normalsize 
A. Einstein\\
June 7, 1928
\\
\vspace{2em} 
\end{center}
\twocolumngrid

\end{appendices}
\setcounter{section}{0}
\setcounter{equation}{0}
\setcounter{page}{1}
\thispagestyle{empty}

Riemann geometry led in general relativity to a physical description of the gravitational field, but does not yield any concepts that can be applied to the electromagnetic field.  For this reason the aim of theoreticians is to find natural generalizations or extensions of Riemann geometry that are richer in content, in hopes of reaching a logical structure that combines all physical field concepts from a single point of view.  Such efforts have led me to a theory which I wish to describe without any attempt at physical interpretation because the naturalness of its concepts lends it a certain interest in its own right.

	Riemannian geometry is characterized by the facts that the infinitesimal neighborhood of every point $P$ has a Euclidian metric, and that the magnitudes of two line elements that belong to the infinitesimal neighborhoods of two finitely distant points $P$ and $Q$ are comparable.  However, the concept of parallelism of these two line elements is missing; for finite regions the concept of direction does not exist.  The theory put forward in the following is characterized by the introduction, in addition to the Riemann metric, of a ``direction," or of equality of direction, or of  ``parallelism" for finite distances.  Correspondingly, new invariants and tensors will appear in addition to those of Riemann geometry.

\section{$n$-Bein and metric}

	At the arbitrary point $P$ of the $n$-dimensional continuum erect an orthogonal $n$-Bein from $n$ unit vectors representing an orthogonal coordinate system.  Let $A_a$ be the components of a line element, or of any other vector, w.r.t. this local system ($n$-Bein). For the description of a finite region introduce furthermore the Gaussian coordinate system $x^\nu$.   Let $A^\nu$ be the $\nu$-components of the vector ($A$) w.r.t. the latter, furthermore let ${h_a}^\nu$ be the  $\nu$ components of the unit vectors that form the $n$-Bein.  
Then\footnote{We use Greek letters for the coordinate indices, Latin letters for Bein indices.}
\begin{equation}
A^\nu=h^\nu_a A_a \; \cdots .
\end{equation}
By inverting (1) and calling $h_{\nu a}$  the normalized sub-determinants (cofactor) of ${h_a}^\nu$ we obtain
\begin{equation}
\tag{1a}
A^a=h^\mu_a A^\mu \; \cdots .
\end{equation}
The magnitude $A$ of the vector ($A$), on account of the Euclidian property of the infinitesimal neighborhoods, is given by 
\begin{equation}
A^2 = \sum A_a^2 = h_{\mu a}h_{\nu a} A^\mu A^\nu \; \cdots .
\end{equation}
The metric tensor components $g_{\mu\nu}$ are given by the formula
\begin{equation}
 g_{\mu \nu} =  h_{\mu a} h_{\nu a}, \; \cdots 
\end{equation}
where, of course, the index $a$ is summed over.  With fixed $a$, the ${h_a}^\nu$   are the components of a contravariant vector.  The following relations also hold:
\begin{equation}
 h_{\mu a} h^\nu_{ a} = \delta^\nu_\mu \; \cdots
\end{equation}
\begin{equation}
 h_{\mu a} h^\mu_{ b} = \delta_{a b}, \; \cdots
\end{equation}
where $\delta =1$ or $\delta =0$ depending on whether the two indices are equal or different.  The correctness of (4) and (5) follows from the above definition of ${h_{\nu a}}$   as normalized subdeterminants of $h_a^\nu$.  The vector character of ${h_{\nu a}}$ follows most easily from the fact that the left hand side, and hence also the right hand side, of (1a) is invariant under arbitrary coordinate transformations for any choice of the vector ($A$).  

	The $n$-Bein field is determined by $n^2$ functions $h_a^\nu$, while the Riemann metric is determined by merely $\frac{n(n +1)}{2}$ quantities $g_{\mu\nu}$.  According to (3) the metric is given by the $n$-Bein field, but not vice versa.

\section{Distant parallelism and rotational invariance}

 	By positing the $n$-Bein field, the existence of the Riemann metric and of distant parallelism are expressed simultaneously.  If ($A$) and ($B$) are two vectors at the points $P$ and $Q$ respectively, which w.r.t. the local $n$-Beins have equal local coordinates ({\it i.e.} $A_a=B_a$) they are to be regarded as equal (on account of (2)) and as ``parallel." 

	If we consider only the essential, {\it i.e.} the objectively meaningful, properties to be the metric and distant parallelism, we recognize that the $n$-Bein field is not yet completely determined by these demands.  Both the metric and distant parallelism remain intact if one replaces the  $n$-Beins of all points of the continuum by others which result from the original ones by a common rotation.  We call this replaceability of the $n$-Bein field Òrotational invarianceÓ and assume:  Only rotationally invariant mathematical relationships can have real meaning.

	Keeping a fixed coordinate system, and given a metric as well as a distant parallelism relationship, the ${h_a}^\mu$ are not  yet fully determined;  a substitution of the ${h_a}^\nu$ is still possible which corresponds to rotational invariance, {\it i.e.} the equation 
\begin{equation}
\tag{6}
A^\ast_a = d_{am} A_m\;\cdots
\end{equation}
where the $d_{am}$ is chosen to be orthogonal and independent of the coordinates.   ($A_a$) is an arbitrary vector w.r.t. the local coordinate system; ($A^\ast_a$) is the same one in terms of the rotated local system.  According to (1a), equation (6) yields
\begin{equation*}
h^\ast_{\mu a} A^\mu = d_{am} h_{\mu m} A^\mu
\end{equation*}
or
\begin{subequations}
\begin{equation}
h^\ast_{\mu a} = d_{am} h_{\mu m} ,\;\cdots
\end{equation}
where
\begin{eqnarray}
d_{am} d_{bm} &=& d_{ma} d_{mb}=\delta_{ab}, \; \cdots  \\
\frac{\partial d_{am} }{\partial x^\nu}&=&0. \; \cdots
\end{eqnarray}
\end{subequations}
The assumption of rotational invariance then requires that equations containing  $h$ are to be regarded as meaningful only if they retain their form when  they are expressed in terms of $h^\ast$ according to (6).  Or:  $n$-Bein fields related by local uniform rotations are equivalent.
	The law of infinitesimal parallel transport of a vector in going from a point ($x^\nu$) to a neighboring point ($x^\nu   +  d x^\nu$)  is evidently characterized by the equation
\begin{equation}
d A_a = 0\; \cdots
\end{equation}
which is to say the equation
\begin{displaymath}
0= d(h_{\mu a } A^\nu)		
=
\frac{\partial h_{\mu a} }{\partial x^\tau} A^\mu dx^\tau +	h_{\mu a} dA^\mu		= 0.
\end{displaymath}
Multiplying by $h_a^\nu$ and using (5), this equation becomes 
\begin{equation}
\tag{7a}
\text{where}
\hspace{0.5in}
\left.
\begin{array}{l}
     dA^\nu=- \Delta^\nu_{\mu\sigma}  A^\mu dx^\tau\\
     {}\\
      \Delta^\nu_{\mu\sigma}  = h^{\nu a} \frac{\partial h_{\mu a} }{\partial x^\sigma} .
\end{array}
\right\}
\end{equation}
This parallel transport law is rotationally invariant and is unsymmetrical with respect to the lower indices of $\Delta^\nu_{\mu\sigma}$.  If the vector ($A$) is moved along a closed path according to this law, it returns to itself; this means that the Riemann tensor $R$, defined in terms of the transport coefficients $\Delta^\nu_{\mu\sigma}$,
\begin{displaymath}
R^i_{k,lm} = - \frac{\partial \Delta^i_{kl} }{\partial x^m}+ \frac{\partial \Delta^i_{km} }{\partial x^l}+\Delta^i_{\alpha l}\Delta^\alpha_{km}-\Delta^i_{\alpha m}\Delta^\alpha_{kl}
\end{displaymath}
will vanish identically because of (7a)---as can be verified easily.

	Besides this parallel transport law there is another (nonintegrable) symmetrical law of transport that belongs to the Riemann metric according to (2) and (3).  It is given by the well-known equations 
\begin{equation}
\left.
\begin{array}{l}
     \overline{d}A^\nu=- \Gamma^\nu_{\mu\sigma}  A^\mu dx^\tau\\
     {}\\
      \Gamma^\nu_{\mu\sigma}  = \frac{1}{2} g^{\nu a}
      \left(
      \frac{\partial g_{\mu \alpha} }{\partial x^\tau}+  \frac{\partial g_{\tau \alpha} }{\partial x^\mu}-  \frac{\partial g_{\mu \sigma} }{\partial x^\alpha}
      \right) .
\end{array}
\right\}
\end{equation}
The $ \Gamma^\nu_{\mu\sigma}$ symbols are given in terms of the $n$-Bein field $h$ according to (3).  It should be noted that 
\begin{equation}
g^{\mu\nu}=h^\mu_a h^\nu_a .\; \cdots
\end{equation}
Equations (4) and (5) imply 
\begin{displaymath}
g^{\mu\lambda} g_{\nu\lambda} =\delta^\mu_\nu
\end{displaymath}
which defines $g^{\mu\nu}$ in terms of $g_{\mu\nu}$.  This law of transport based on the metric is of course also rotationally invariant in the sense defined above.

 \section{Invariants and covariants}

	In the manifold we have been studying, there exist, in addition to the tensors and invariants of Riemann geometry, which contain the quantities $h$ only in the combinations given by (3), further tensors and invariants, of which we want to consider only the simplest.  
	
		Starting from a vector ($A^\nu$) at the point ($x^\nu$), the two transports $d$ and $\bar d$ to the neighboring point ($x^\nu + dx^\nu$) result in the two vectors
\begin{equation*}
A^\nu +dA^\nu
\end{equation*}

and
\begin{equation*}
A^\nu + \overline{d} A^\nu.
\end{equation*}
The difference 
\begin{equation*}
dA^\nu -  \overline{d} A^\nu = (\Gamma^\nu_{\alpha\beta} - \Delta^\nu_{\alpha\beta}) A^\alpha dx^\beta
\end{equation*}
is also a vector.  Hence 
\begin{equation*}
\Gamma^\nu_{\alpha\beta} - \Delta^\nu_{\alpha\beta}
\end{equation*}
is a tensor, and so is its antisymmetric part
\begin{displaymath}
\frac{1}{2}(\Delta^\nu_{\alpha\beta} - \Delta^\nu_{\beta\alpha}) = \Lambda^\nu_{\alpha\beta}.\;\cdots
\end{displaymath}
The fundamental meaning of this tensor in the theory here developed emerges from the following:  If this tensor vanishes, the continuum is Euclidian.   For if
\begin{equation}
		0 = 2\Lambda^\nu_{\alpha\beta} = h^a
		      \left(
      \frac{\partial h_{\alpha a} }{\partial x^\beta}+  \frac{\partial h_{\beta a} }{\partial x^\alpha}
            \right),
\end{equation}
then multiplication by $h_{\nu b}$  yields
\begin{displaymath}
 		0 =      \frac{\partial h_{\alpha b} }{\partial x^\beta}+  \frac{\partial h_{\beta b} }{\partial x^\alpha}.
\end{displaymath}
We can therefore put 
\begin{displaymath}
h_{\alpha b} = \frac{\partial\Psi_b}{\partial x^\alpha}.
\end{displaymath}
The field is therefore derivable from $n$ scalars $\Psi_b$.  We choose the coordinates according to the equation
\begin{equation}
\Psi_b = x^b .
\end{equation}
Then, according to (7a) all $\Delta^\nu_{\alpha\beta}$ vanish, and the $h_{\mu a}$ as well as the $g_{\mu\nu}$ are constant.

	Since the tensor $\Lambda^\nu_{\alpha\beta}$ is evidently also formally the simplest one allowed by our theory, the simplest characterization of the continuum will be tied to $\Lambda^\nu_{\alpha\beta}$, not to the more complicated Riemann curvature tensor.  The simplest forms that can come into play here are the vector 
\begin{equation*}
\Lambda^\alpha_{\mu\alpha}
\end{equation*}
as well as the invariants
\begin{equation*}
g^{\mu\nu} \Lambda^\alpha_{\mu\beta} \Lambda^\beta_{\nu\alpha}  
\qquad \text{and} \qquad  g_{\mu\nu} g^{\alpha\sigma}  g^{\beta\tau}  \Lambda^\mu_{\alpha\beta}\Lambda^\nu_{\sigma\tau} .
\end{equation*}
From one of the latter (or from linear combinations) an invariant integral $J$ can be constructed by multiplication with the invariant volume element
\begin{equation*}
h \; d\tau ,
\end{equation*}
where $h$ is the determinant of $|h_{\mu a}|$,  and $d\tau$ is the product $dx_1 \dots dx_n$.  The assumption
\begin{equation*}
\delta J = 0
\end{equation*}
yields 16 differential equations for the 16 values of $h_{\mu a}$.

	Whether one can get physically meaningful laws in this way will be investigated later.

	It is helpful to compare Weyl's modification of Riemann's theory with the theory developed here:
\begin{quote}
  WEYL:  Comparison neither of distant vector magnitudes nor of directions;
\end{quote} 
\begin{quote}
  RIEMANN: Comparison of distant vector magnitudes, but not of distant directions;
\end{quote} 
\begin{quote}
  THIS THEORY:  Comparison of distant vector magnitude and directions. 
\end{quote}


\onecolumngrid
\newpage
\section*{Einstein's 1928 manuscript on unification of gravity and electromagnetism}
\thispagestyle{empty}
\newpage
\begin{center}
\mbox{
{\bf\large New possibility for a unified field theory of gravity and electricity}
}
\\
\vspace{1em}
\normalsize 
A. Einstein\\
June 14, 1928
\\
\vspace{2em} 
\end{center}
\twocolumngrid
\setcounter{section}{0}
\setcounter{equation}{0}
\setcounter{page}{1}
\thispagestyle{empty}

A few days ago I explained in a short paper in these Reports how it is possible to use an n-Bein-Field to formulate a geometric theory based on the fundamental concepts of the Riemann metric and distant parallelism.  At the time I left open the question whether this theory could serve to represent physical relationships.  Since then I have discovered that this theory---at least in first approximation---yields the field equations of gravity and electromagnetism very simply and naturally.  It is therefore conceivable that this theory will replace the original version of the theory of relativity.

	The introduction of distant parallelism implies that according to this theory there is something like a straight line, {\it i.e.} a line whose elements are all parallel to each other; of course such a line is in no way identical to a geodesic.  Furthermore, in contrast to the usual general theory of relativity, there is the concept of relative rest of two mass points (parallelism of two line elements which belong to two different worldlines.)

	In order for the general theory to be useful immediately as field theory one must assume the following:

\begin{enumerate}
  \item The number of dimensions is 4 ($n=4$).
  \item The fourth local component $A_a$ ($a=4$) of a vector is pure imaginary, and hence so are the components of the four legs of the Vier-Bein, the quantities ${h^\mu}_4$  and $h_{\mu 4}$.\footnote{Instead one could also define the square of the magnitude of the local vector $A$ to be $A_1^2 + A_2^2+ A_3^2 - A_4^2$ and introduce Lorentz transformations instead of rotations of the local n-Bein.  In that case all the $h$'s would be real, but the immediate connection with the general theory would be lost.}
\end{enumerate}
The coefficients $g_{\mu\nu}\; (= h_{\mu\alpha} h_{\nu\alpha})$ of course all become real.  Accordingly, we choose the square of the magnitude of a timelike vector to be negative. 

\section{The underlying field equation}

Let the variation of a Hamiltonian integral vanish for variations of the field potentials $h_{\mu\alpha}$  (or $h^\mu_\alpha$ ) that vanish on the boundary of a domain:
\begin{subequations}
\begin{equation}
\tag{1}
\delta \left\{\int \mathfrak{H} d\tau\right\} = 0. \;\cdots
\end{equation}%
\begin{equation}
\label{ }
 \mathfrak{H} = h\, g^{\mu\nu} \; {\Lambda_\mu}^\alpha_\beta \; {\Lambda_\nu}^\beta_\alpha, \;\cdots
\end{equation}
\end{subequations}
where the quantities $h \; (= \det h_{\mu\alpha} )$, $g^{\mu\nu}$, and $\Lambda^\alpha_{\mu\nu}$ are defined in (9) and (10) of the previous paper.  

Let the $h$ field describe the electrical and the gravitational field simultaneously.  A ``purely gravitational field'' results when equation (1) is fulfilled and, in addition, 
\begin{equation}
\phi_\mu = {\Lambda_\mu}^\alpha_\alpha\;\cdots
\end{equation}
vanish, which represents a covariant and rotationally invariant subsidiary condition.\footnote{Here there remains a certain ambiguity of interpretation, because one could also characterize the pure gravitational field by the vanishing of  $\frac{\partial\phi_\mu}{\partial x_\nu}- \frac{\partial\phi_\nu}{\partial x_\mu}$.}

\section{The field equation in the first approximation}

If the manifold is the Minkowski world of special relativity, one can choose the coordinates in such a way that $h_{11} = h_{22} = h_{33} = 1 , h_{44} = j\; ( =  \sqrt{-1})$, and that all other $h$'s vanish.  This set of values is somewhat inconvenient for calculation.  For that reason we prefer to choose the $x_4$ coordinate in this \S ~to be pure imaginary; in that case the Minkowski world (absence of any field when the coordinates are chosen appropriately) can be described by 
\begin{equation}
h_{\mu a}=\delta_{\mu a}\;\cdots
\end{equation}
The case of infinitely weak fields can be represented suitably by
\begin{equation}
h_{\mu a}=\delta_{\mu a} + k_{\mu\alpha}\;\cdots
\end{equation}
where the $k_{\mu\alpha}$ are small quantities of first order. Neglecting quantities of third or higher order we must replace (1a) by (1b), considering (10) and (7a) of the previous paper:  
\begin{equation}
\tag{1b}
 \mathfrak{H}  = -\frac{1}{4}
 \left(\frac{\partial k_{\mu\alpha}}{\partial x_\beta}- \frac{\partial  k_{\beta\alpha}}{\partial x_\mu}\right)
\left(\frac{\partial k_{\mu\beta}}{\partial x_\alpha}- \frac{\partial  k_{\alpha \beta}}{\partial x_\mu}\right).\;\cdots
\end{equation}
After variation one obtains the field equations in the first approximation 
\begin{equation}
\frac{\partial^2 k_{\beta\alpha}}{\partial x_\mu^2}-\frac{\partial^2 k_{\mu\alpha}}{\partial x_\mu\partial x_\beta}+\frac{\partial^2 k_{\alpha \mu}}{\partial x_\beta\partial x_\mu}-\frac{\partial^2 k_{\beta\mu}}{\partial x_\mu\partial x_\alpha} =0. \; \cdots
\end{equation}
These are 16 equations\footnote{On account of general covariance there are of course four identities among the field equations.  In the first approximation considered here, this is expressed by the fact that the divergence of the left side of (5) with respect to the index $\alpha$ vanishes identically.} for the 16 components $k_{\alpha\beta}$.
Our task now is to see whether this system of equations contains the known laws of the gravitational and electromagnetic fields.  For this purpose we must introduce $g_{\alpha\beta}$ and $\phi_\alpha$ in (5) in place of $k_{\alpha\beta}$.  We must put
\begin{displaymath}
g_{\alpha\beta}=h_{\alpha a}h_{\beta a} = (\delta_{\alpha a} + k_{\alpha a})(\delta_{\beta a} + k_{\beta a}).
\end{displaymath}
Or, exact to first order, 
\begin{equation}
g_{\alpha\beta}-\delta_{\alpha\beta}= \overline{g_{\alpha\beta}}=k_{\alpha\beta}+k_{\beta\alpha}.\; \cdots
\end{equation}
From (2) one also gets the quantities to first order exactly
\begin{equation}
\tag{2a}
2 \phi_\alpha=\frac{\partial k_{\alpha \mu}}{\partial x_ \mu}- \frac{\partial  k_{\mu \mu}}{\partial x_\alpha}
\end{equation}
By exchanging $\alpha$ and $\beta$ in (5) and adding to (5) one gets 
\begin{displaymath}
           \frac{\partial^2 \overline{g_{\alpha\beta}}}{\partial x_\mu^2}
           -\frac{\partial^2 k_{\mu\alpha}}{\partial x_\mu\partial x_\beta}
           -\frac{\partial^2 k_{\mu\beta}}{\partial x_\mu\partial x_\alpha}          								= 0 .
\end{displaymath}
If one adds to this equation the two following equations which follow from (2a)
\begin{eqnarray*}
    -\frac{\partial^2 k_{\alpha\mu}}{\partial x_\mu\partial x_\beta}
           +\frac{\partial^2 k_{\mu \mu}}{\partial x_\beta\partial x_\alpha} 
            & = & 
            -2 \frac{\partial\phi_\alpha}{\partial x_\beta}\\
    -\frac{\partial^2 k_{\beta\mu}}{\partial x_\mu\partial x_\alpha}
           +\frac{\partial^2 k_{\mu\mu}}{\partial x_\alpha\partial x_\beta} 
            & = &   -2 \frac{\partial\phi_\beta}{\partial x_\alpha} ,
\end{eqnarray*}
then one obtains, in view of (6), 
\begin{eqnarray}
\nonumber
\frac{1}{2}\left(
 -\frac{\partial^2 \overline{g_{\alpha\beta}}}{\partial x_\mu^2}
          +\frac{\partial^2 \overline{g_{\mu\alpha}}}{\partial x_\mu\partial x_\beta}
          +\frac{\partial^2 \overline{g_{\mu\beta}}}{\partial x_\mu\partial x_\alpha}
          - \frac{\partial^2 \overline{g_{\mu\mu}}}{\partial x_\alpha\partial x_\beta}
\right)
\\
=\frac{\partial\phi_\alpha}{\partial x_\beta}
+
 \frac{\partial\phi_\beta}{\partial x_\alpha}.\;\cdots
\end{eqnarray}
The case of vanishing electromagnetic fields is characterized by the vanishing of  $\phi_\alpha$.  In that case, (7) agrees to first order with the equation of General Relativity
\begin{equation*}
R_{\alpha\beta}=0
\end{equation*}
(where $R_{\alpha\beta}$ is the once reduced Riemann tensor.)  Thus it is proved that our new theory correctly reproduces the law of the pure gravitational field in the first approximation.

	By differentiating (2a) with respect to $x_\alpha$ and taking into account the equation obtained from (5) by reducing with respect to $\alpha$ and $\beta$ one obtains 
\begin{equation}
\frac{\partial\phi_\alpha}{\partial x_\alpha}=0.\;\cdots
\end{equation}
Noting that the left side $L_{\alpha\beta}$ of (7) obeys the identity 
\begin{equation*}
\frac{\partial}{\partial x_\beta}
\left(
L_{\alpha\beta}-\frac{1}{2}\delta_{\alpha\beta}L_{\sigma\sigma}
\right)=0,
\end{equation*}
we find from (7) that
\begin{displaymath}
\frac{\partial^2 \phi_\alpha}{\partial x_\beta^2}
          +\frac{\partial^2\phi_\beta}{\partial x_\alpha\partial x_\beta}
          -\frac{\partial}{\partial x_\alpha}
          \left(
          \frac{\partial\phi_\sigma}{\partial x_\sigma}
          \right)
          =0
\end{displaymath}
or
\begin{equation}
\frac{\partial^2\phi_\alpha}{\partial x_\beta^2}=0.\;\cdots
\end{equation}
Equations (8) and (9) together are known to be equivalent to Maxwell's equations for empty space.  The new theory therefore yields Maxwell's equations in first approximation.

 	The separation of the gravitational field from the electromagnetic field appears artificial according to this theory.  And it is clear that equations (5) imply more than (7), (8), and (9) together.  Furthermore, it is remarkable that in this theory the electric field does not enter the field equations quadratically. 

Note added in proof:  One obtains very similar results by starting with the Hamiltonian
\begin{equation*}
 \mathfrak{H} =  h g_{\mu\nu} g^{\alpha\sigma} g^{\beta\tau} {\Lambda^\mu_\alpha}_\beta  {\Lambda^\nu_\sigma}_\tau.
\end{equation*}
There is therefore at this time a certain uncertainty with respect to the choice of  $\mathfrak{H}$.

\end{document}